# Commissioning of the NUCLEUS Experiment at the Technical University of Munich

H. Abele,[1] G. Angloher,[2] B. Arnold,[3] M. Atzori Corona,[4] A. Bento,[2,*] E. Bossio,[5,†] F. Buchsteiner,[3] J. Burkhart,[3] F. Cappella,[6] M. Cappelli,[7,6] N. Casali,[6] R. Cerulli,[4] A. Cruciani,[6] G. Del Castello,[6] M. del Gallo Roccagiovine,[7,6] S. Dorer,[1] A. Erhart,[8] M. Friedl,[3] S. Fichtinger,[3] V. M. Ghete,[3] M. Giammei,[9,4] C. Goupy,[5,‡] D. Hauff,[2,8] F. Jeanneau,[5] E. Jericha,[1] M. Kaznacheeva,[8] H. Kluck,[3] A. Langenkämper,[2] T. Lasserre,[5,8,‡] D. Lhuillier,[5] M. Mancuso,[2] R. Martin,[5,1] B. Mauri,[2] A. Mazzolari,[10,11] L. McCallin,[5] H. Neyrial,[5] C. Nones,[5] L. Oberauer,[8] T. Ortmann,[8,§] L. Peters,[8,5,‡] F. Petricca,[2] W. Potzel,[8] F. Pröbst,[2] F. Pucci,[2] F. Reindl,[3,1] M. Romagnoni,[10,11] J. Rothe,[8] N. Schermer,[8] J. Schieck,[3,1] S. Schönert,[8] C. Schwertner,[3,1] L. Scola,[5] G. Soum-Sidikov,[5] L. Stodolsky,[8] R. Strauss,[8] R. Thalmeier,[3] C. Tomei,[6] M. Vignati,[7,6] M. Vivier,[5] and A. Wex[8]

(NUCLEUS Collaboration)

[1]*Atominstitut, Technische Universität Wien, Stadionallee 2, Wien, A-1020, Austria*
[2]*Max-Planck-Institut für Physik, Boltzmannstraße 8, Garching, D-85748, Germany*
[3]*Institut für Hochenergiephysik der Österreichischen Akademie der Wissenschaften, Dominikanerbastei 16, Wien, A-1010, Austria*
[4]*Istituto Nazionale di Fisica Nucleare—Sezione di Roma "Tor Vergata", Via della Ricerca Scientifica 1, Roma, I-00133, Italy*
[5]*IRFU, CEA, Université Paris-Saclay, Bâtiment 141, Gif-sur-Yvette, F-91191, France*
[6]*Istituto Nazionale di Fisica Nucleare—Sezione di Roma, Piazzale Aldo Moro 2, Roma, I-00185, Italy*
[7]*Dipartimento di Fisica, Sapienza Università di Roma, Piazzale Aldo Moro 5, Roma, I-00185, Italy*
[8]*Physik-Department, TUM School of Natural Sciences, Technische Universität München, James-Franck-Straße 1, Garching, D-85748, Germany*
[9]*Dipartimento di Fisica, Università di Roma "Tor Vergata", Via della Ricerca Scientifica 1, Roma, I-00133, Italy*
[10]*Istituto Nazionale di Fisica Nucleare—Sezione di Ferrara, Via Giuseppe Saragat 1c, Ferrara, I-44122, Italy*
[11]*Dipartimento di Fisica, Università di Ferrara, Via Giuseppe Saragat 1, Ferrara, I-44122, Italy*

The NUCLEUS experiment aims to detect coherent elastic neutrino-nucleus scattering of reactor antineutrinos on $CaWO_4$ targets in the fully coherent regime, using gram-scale cryogenic calorimeters. The experimental apparatus will be installed at the Chooz nuclear power plant in France, in the vicinity of two 4.25 $GW_{th}$ reactor cores. This work presents results from the commissioning of an essential version of the experiment at the shallow Underground Laboratory of the Technical University of Munich. For the first time, two cryogenic target detectors were tested alongside active and passive shielding systems. Over a period of two months, all detector subsystems were operated with stable performance. Background measurements were conducted, providing important benchmarks for the modeling of background sources at the reactor site. Finally, we present ongoing efforts to upgrade the detector systems in preparation for a technical run at Chooz in 2026, and highlight the remaining challenges to achieving neutrino detection.

---

*Also at: LIBPhys-UC, Departamento de Física, Universidade de Coimbra, Rua Larga 3004-516, Coimbra, P3004-516, Portugal.
†Contact author: elisabetta.bossio@cea.fr
‡Permanent address: Max-Planck-Institut für Kernphysik, Saupfercheckweg 1, Heidelberg, D-69117, Germany.
§Permanent address: Max-Planck-Institut für Physik, Boltzmannstraße 8, Garching, D-85748, Germany.

## I. INTRODUCTION

Coherent elastic neutrino-nucleus scattering (CEνNS) is a fundamental process predicted over four decades ago [1,2] and first observed in 2017 by the COHERENT Collaboration using neutrinos from a spallation source scattering off a CsI target [3]. Subsequent measurements in Ar [4] and Ge [5] from the same collaboration have firmly established CEνNS as an observable within reach of experimental techniques, opening new paths in neutrino physics and searches for physics beyond the standard model [6,7].

Nuclear reactors, the most intense terrestrial sources of antineutrinos used for research, offer promising opportunities for CEνNS detection. A high-statistics and precise measurement of CEνNS at reactors would open a broad physics program. At reactor neutrino energies, the interaction occurs in the fully coherent regime, rendering the cross section independent of nuclear form factors. This enables clean tests of the standard model, such as measurements of the weak mixing angle at low-momentum transfer [8], and searches for new physics, including nonstandard neutrino interactions and new light mediator particles [9]. It also opens the door to applications ranging from astrophysics to reactor monitoring [10,11]. After years of extensive efforts by numerous reactor neutrino experiments [12–21], the first detection of CEνNS from reactor antineutrinos was recently reported by the CONUS+ experiment on a Ge target [22]. This milestone demonstrates the feasibility of reactor-based CEνNS detection, yet significant challenges remain for precision measurements. The main difficulties are posed by the ultralow nuclear recoil energies induced by reactor neutrinos, typically below 1 keV, the unfavorable background conditions commonly encountered at reactor sites, and the need to address both challenges with a sufficient target mass.

The NUCLEUS experiment aims to overcome these obstacles by employing gram-scale cryogenic calorimeters that can reach extremely low thresholds on nuclear recoils, around 20 eV [23,24]. The calorimeters are integrated into a compact veto system, incorporating both active and passive shielding, designed to achieve a CEνNS-signal-to-particle background ratio of approximately 1 in the energy region of interest (20–100 eV) [25], where *particle background* denotes backgrounds from known particle interactions, in contrast to unexplained contributions, such as the low-energy excess (LEE) [26]. The experiment will be deployed at the Chooz nuclear power plant in France, at the so-called very near site (VNS), located 102 m and 72 m from two 4.25 GW$_{th}$ reactor cores [27]. The same power plant also previously hosted the Double Chooz experiment [28], albeit at a different site. The average antineutrino flux expected at the VNS is $1.7 \times 10^{12}$ cm$^{-2}$ s$^{-1}$, calculated considering an average reactor operating cycle of 80% and using the reactor neutrino flux model from Ref. [29].

In order to prepare for installation at Chooz, a simplified version of the NUCLEUS experiment was commissioned at the shallow Underground Laboratory (UGL) of the Technical University of Munich (TUM). All available NUCLEUS subsystems were operated together for the first time, but the commissioned setup still differed from the final configuration at Chooz in several respects: only two out of 18 target detectors were installed, the inner active veto was absent, only one out of the six cryogenic outer veto (COV) crystals was installed, one layer of passive shielding inside the cryostat was missing, and a temporary two-channel data acquisition (DAQ) system was in use. Despite these differences, this commissioning phase marked an important milestone for the collaboration: an eight-week continuous data acquisition run was completed, demonstrating the stable operation of the cryogenic detectors in a dry cryostat, cooled by a pulse-tube refrigerator, over month-long periods. Additional technical achievements included the successful operation of the detectors in coincidence with the veto systems [30,31] and the commissioning of the LED optical calibration system [32]. These steps are critical for ensuring the readiness of the experiment for deployment at the reactor site.

From a physics perspective, the commissioning run was designed to measure the background in the CEνNS region of interest in preparation for the forthcoming neutrino measurement at the reactor. While this objective remains, the data revealed that a previously identified LEE dominates the CEνNS region of interest, precluding a direct measurement of the particle background there at this stage. The full background characterization will be completed in future work. This article instead presents key results on the particle background in the adjacent keV energy range, providing essential input for validating the shielding strategy and benchmarking the simulation framework. This analysis is particularly relevant since both experimental sites are at shallow depth. In parallel, dedicated studies of the LEE were conducted, laying the groundwork for a dedicated analysis to be presented in a forthcoming publication and informing the design of the cryogenic detectors to be deployed at Chooz to optimally suppress this background.

This paper presents results from the first commissioning run of the NUCLEUS experiment at TUM. Section II describes the experimental setup and outlines the key differences compared to the final configuration planned at the Chooz site. Section III presents the performance and stability of all detector subsystems. Measurements of the particle background, along with comparisons to Geant4-based simulations in the keV energy range, are discussed in Sec. IV. Section V details the ongoing efforts to upgrade the experimental setup in preparation for its relocation to Chooz, where a first technical run at the reactor site is planned for 2026, and discusses the remaining challenges to be addressed in achieving neutrino detection. A summary and outlook are given in Sec. VI.

## II. EXPERIMENTAL SETUP

The experimental setup was commissioned in a shallow underground laboratory at TUM, which provides an overburden of approximately 10 m water equivalent (w.e.) [33]. For comparison, the VNS at Chooz offers an even shallower overburden of only about 3 m w.e. A schematic view of the experimental setup is shown in Fig. 1, where all key components are labeled and differences to the final NUCLEUS setup at Chooz are highlighted. The following sections describe each component in detail.

*a. Cryostat and vibration decoupling system* The experiment operates a BlueFors LD400 cryogen-free dilution refrigerator that reaches a base temperature below 8 mK, with a cooling power of 17.5 $\mu$W at 20 mK [34]. Cooling down to 4 K is achieved using a Cryomech PT-415 pulse tube operating at 1.4 Hz. In order to minimize transmission of vibrations, the detectors are mechanically decoupled from the cryostat through a 1.8-meter-long cryogenic spring pendulum attached to an autonomous reference frame, a system described in Ref. [35].

*b. Passive shielding* The passive shielding consists of layers external and internal to the cryostat, in order to provide a nearly complete $4\pi$ coverage of the cryogenic detectors [25]. Both internal and external shielding components are made of the same materials with identical thicknesses, and the inner layers were vertically aligned with the external ones in order to minimize the solid angle of unavoidable gaps caused by the cryostat mechanics. The shielding structure includes a 5 cm layer of low-radioactivity lead to suppress environmental gamma radiation, combined with a 20 cm inner layer of 5% borated polyethylene (PE) for neutron attenuation. The external shielding measures $93 \times 93 \times 86$ cm³ with a cylindrical opening (430 mm in diameter) from the top to accommodate the cryostat. A movable steel frame mounted on rails supports the rectangular PE/lead shielding and allows access to the cryostat when needed. The internal shielding, positioned directly above the cryogenic target detectors, has a cylindrical shape with a diameter of 297 mm and is mechanically secured to the cryostat using a bayonet mount. It is thermally connected to the still stage of the cryostat via a cold finger made of NOSV copper, a high-conductivity copper alloy produced by Aurubis [36] and commonly used in cryogenic applications, and thermalized at 800 mK. Several interleaved copper strips ensure uniform thermalization across the internal shielding structure. It is important to note that the passive shielding described above—and commissioned in this work—is identical to the configuration planned for Chooz, with just one exception: an additional boron carbide ($B_4C$) layer surrounding the target detectors. This extra layer will be crucial at Chooz for suppressing cosmic-ray-induced neutron background, whereas in the UGL the hadronic component of cosmic radiation is already sufficiently suppressed due to the additional overburden of the experimental site.

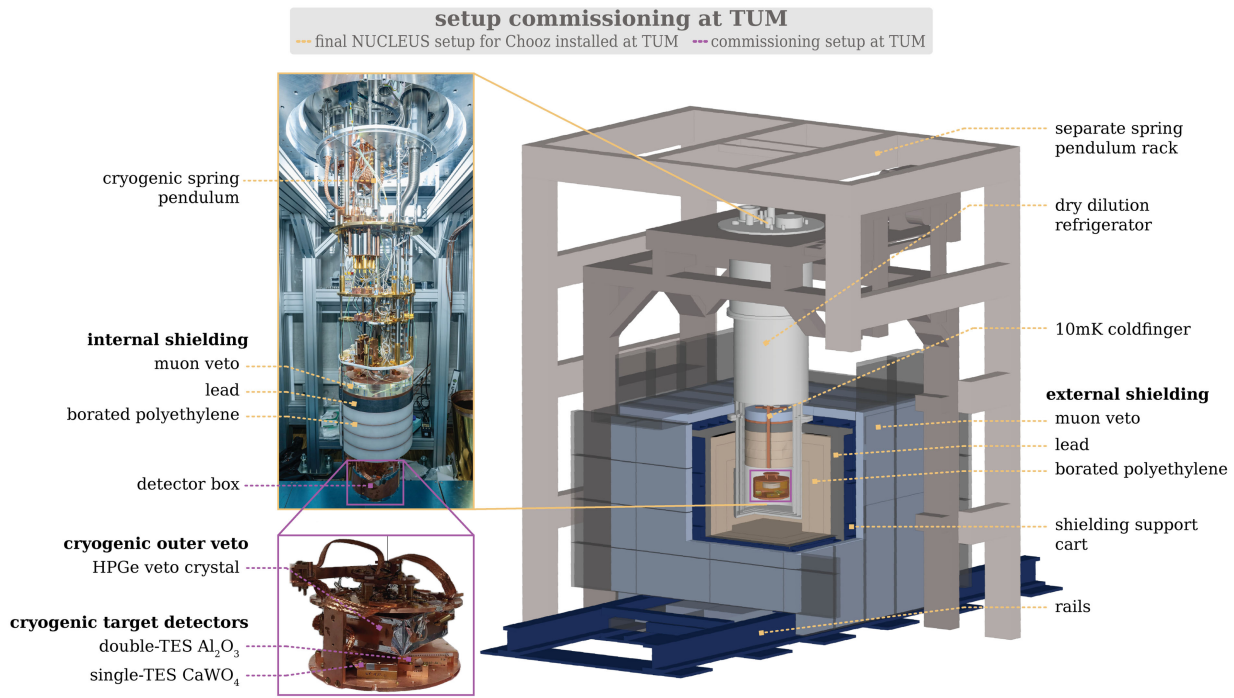

FIG. 1.   Schematic and photographic representation of the NUCLEUS commissioning setup at TUM. On the right, a 3D model of the experiment highlights the key structural components. The upper left insets provide photos of the cryogenic system, while the bottom left inset focuses on the inside of the detector box with the cryogenic outer veto and the two cryogenic target detectors. The latter is the main difference to the final NUCLEUS setup at Chooz, which will include 18 target detectors arranged in an instrumented silicon holder, six high-purity germanium detectors forming the cryogenic outer veto, and an additional boron carbide layer around it.

*c. External and internal muon veto* In order to efficiently tag muons that could produce backgrounds when interacting in the surrounding materials, the experiment employs a muon veto (MV) system [37]. The external MV system consists of 28 modules arranged in a cubic configuration around the cryostat, attached to the outer steel structure enclosing the passive shielding. Each module consists of a plastic scintillator panel and is read out by one or two silicon photomultipliers (SiPMs) from KETEK (model PE3325-WB-TIA-SP). The system is configured across the six faces of the cube: the top face contains eight panels, the bottom face contains four panels, and the four side faces contain a total of 16 panels. On two opposite side faces, the eight panels are mounted with their long axis oriented horizontally, while on the other two opposite sides, the eight panels are mounted with their long axis oriented vertically. To ensure improved light collection, the eight horizontally mounted side panels and all four bottom panels are equipped with a double SiPM readout. In contrast, the eight vertically mounted side panels and all eight top panels are read out by a single SiPM each, due to space constraints. This configuration results in a total of 40 readout channels. Wavelength-shifting optical fibers of type BC-91A from Saint-Gobain, embedded in grooves along each panel, ensure compact readout and efficient light collection. In addition to the external muon veto, a disk-shaped cryogenic muon veto is located inside the cryostat, directly above the internal passive shielding [38]. This veto follows the same SiPM-based readout approach, with fibers routed beneath the 300 K cryostat plate up to a preamplified and gain-stabilized SiPM module from KETEK (model PE3325-WB-TIA-TP) used for signal readout.

*d. Cryogenic outer veto* A core component in the background mitigation strategy of NUCLEUS constitutes the COV. The COV designed for the final NUCLEUS setup at Chooz consists of an arrangement of six high-purity germanium detectors operated at around 10 mK and read out through the ionization channel, which provides a nearly $4\pi$ coverage around the target detectors [30,31]. In the scope of the presented commissioning, only one of the six crystals (described below) was installed directly above the target detectors with reduced coverage. Thanks to its active readout, the COV serves as an efficient tool for identifying and rejecting background events, particularly those induced by environmental gamma radiation. This significantly reduces the need for thick passive lead shielding. At the same time the COV is able to detect muons, thus complementing the muon veto, and features also some neutron detection capability. The installed crystal features a cylindrical geometry with 100 mm diameter and 25 mm height, and a mass of 1 kg. Aluminum electrodes are evaporated on the top and bottom surfaces in plain planar geometry. Readout and thermal link are achieved via copper-Kapton bond pads, which are glued directly onto the crystal [30]. The charge signal of the germanium detector is read out through one of the two electrodes (while the other one is grounded). It is preamplified by a junction field effect transistor (JFET) and then further amplified by an AmpTek A250 charge amplifier [39], following the Edelweiss-II readout scheme [40]. For optimal noise reduction, the JFET is located at the 40 K stage. The RC feedback loop required by the charge amplifier, and the 2 nF decoupling capacitance and the 1 GΩ polarizing resistance, are located at the 4 K stage of the cryostat to reduce their Johnson noise. In order to minimize parasitic capacitance, carbon-coated low-noise and low-capacity constantan coaxial cables by Axon are used inside the cryostat [41]. The output voltage is then filtered and postamplified, to increase the signal-to-noise ratio of the detector amplification chain and to match the input impedance of the data acquisition system.

*e. Target detectors* The core of the NUCLEUS experiment consists of cryogenic calorimeters equipped with tungsten transition-edge sensors (W-TESs) [23]. Two types of detectors were used in the commissioning run: a CaWO$_4$ single-TES detector, shaped as a $5 \times 5 \times 5$ mm$^3$ cube with a mass of 0.76 g and a TES transition temperature of approximately 15.0 mK, and an Al$_2$O$_3$ double-TES detector, measuring $5 \times 5 \times 7.5$ mm$^3$, with a mass of 0.75 g and transition temperatures of the two TESs of 15.1 mK and 15.6 mK, respectively. For stable operation, the TESs were biased with a constant current of 2 $\mu$A, and their operating resistance was regulated by a resistive heater formed by a thin gold film adjacent to the TES structure. Each detector was fully enclosed within a NOSV copper housing to shield it from thermal radiation. The detectors were secured with bronze clamps, while sapphire spheres provided thermal and electrical isolation between the detector, holder, and clamps. The signal readout was facilitated by a customized printed circuit board featuring copper-Kapton-copper traces, which connected the TES wire bonds to external pin connectors for stable data acquisition. All components underwent a thorough cleaning procedure involving isopropyl alcohol and acetone, followed by citric acid etching in a hydrogen peroxide solution to remove oxidation layers and enhance thermal conductivity. Each detector holder was individually mounted inside a copper detector box, with copper strips serving as additional thermal links between the holder and the box. For optical calibration, each holder includes a dedicated hole allowing for the insertion of light fibers with a 120 $\mu$m diameter. Fibers are pulsed at room temperature with a controlled LED [32]. The TES signals are read out using superconducting quantum interference devices (SQUIDs), which are highly sensitive magnetometers that measure the change in resistance of the TES film caused by particle interactions in the target detector. This resistance change alters the current in the readout circuit, resulting in a change in the magnetic flux, which the SQUID then converts into a measurable voltage signal [42].

*f. Data acquisition* The experiment relies on two DAQ systems: a Struck acquisition system—SIS3316 with 16 bit resolution and 125 MHz sampling frequency [43,44]—for the muon veto, and the versatile data acquisition (VDAQ) system [45] for both the COV and the cryogenic target detectors. The VDAQ was developed by members of the NUCLEUS Collaboration specifically for cryogenic experiments. It integrates heater/bias electronics for the optimization and stabilization of TES detectors, as well as digitizing electronics for data recording at sampling frequencies up to 1 MHz. The current setup employs a preliminary version of the VDAQ with two channels only (VDAQ2). The final configuration at Chooz will use an upgraded system with enough channels to accommodate the 18 cryogenic target detectors, the six COV detectors, and the four inner active veto detectors.

Precise time synchronization between the two DAQ systems is achieved using a synchronization module that generates the Struck FADC clock signals. The VDAQ2 system functions as the master clock, periodically sending a synchronization pulse that resets the clocks of the Struck modules to zero, ensuring alignment of the time stamps.

## III. DETECTOR PERFORMANCE AND STABILITY

### A. Measurement overview

The commissioning run lasted eight weeks—from August 1 to September 27, 2024—during which data were collected under operational constraints due to the availability of only two readout channels in the VDAQ2. Since both channels were required for the $Al_2O_3$ detector's double-TES readout system, data acquisition alternated weekly between the $Al_2O_3$ and $CaWO_4$ detectors. Regular LED runs were performed to calibrate the detectors and monitor their stability.

Figure 2 illustrates the cumulative operational time of each detector. In total, approximately three weeks of high-quality data were obtained per target detector, with 571.9 hours recorded for $Al_2O_3$ and 562.6 hours for $CaWO_4$ over a total measurement period of 1135.2 hours, corresponding to duty cycles of 50.4% and 49.6%, respectively, and an overall duty cycle of 83%. Data loss resulted from synchronization issues between the two DAQ systems and human errors, all of which are expected to be mitigated in the upgraded setup at Chooz through a enhanced DAQ system and improved slow-control infrastructure.

The MV remained fully operational throughout the measurement period, accumulating 1135.2 hours. In contrast, the COV was used periodically, as it could only be read out in coincidence with the $CaWO_4$ detector due to the limited VDAQ2 readout channels, for a total of 369.9 hours (see Fig. 2). While the limited rejection power of a single germanium COV crystal makes it less critical for background suppression in the target detectors, its primary value lies in providing an *in situ* measurement of the background environment near the detectors and characterization of the MV rejection efficiency. Importantly, this measurement requires only a few days of data acquisition to reach the required precision.

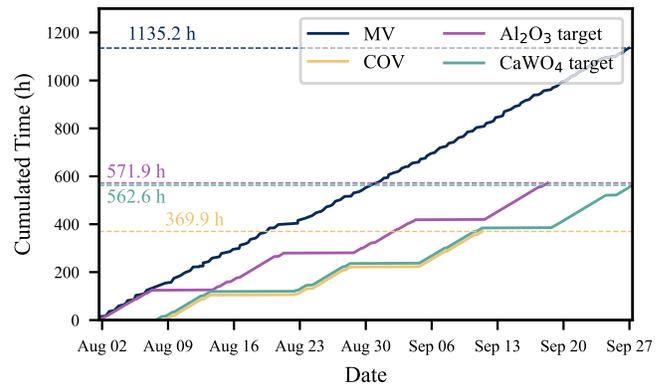

FIG. 2. Cumulative operational time for all detectors during the eight-week commissioning run. The $Al_2O_3$ and $CaWO_4$ detectors were alternated weekly, with the MV running continuously and the COV operating only with $CaWO_4$. The total cumulated exposures are indicated.

Following the main eight-week commissioning run, a short follow-up measurement was conducted using a reduced shielding configuration. During this period, data were acquired from the COV and the $CaWO_4$ target detectors. The primary objective of this measurement was to refine and benchmark the background simulations in a configuration more sensitive to external backgrounds, as will be explained in Sec. IV.

### B. The muon veto detectors

Since the NUCLEUS experiment will be located at shallow depth, cosmic radiation is expected to significantly contribute to the total background [25]. Efficient suppression of this particle background component is essential to achieve the experiment's low-background goals. The performance of the MV system was extensively investigated during the commissioning run, particularly in coincidence with the COV and cryogenic target detectors.

The output signal of the 41 MV channels was sampled at a frequency of 125 MHz. Upon triggering, a waveform of 100 samples (60 pretrigger and 40 post-trigger) is stored. The trigger time and the pulse charge are calculated online, with the latter determined by integrating the post-trigger samples and subtracting the pretrigger baseline. When one channel triggers, signals from all other channels are also recorded as part of the same event, enabling the system to register a muon crossing multiple panels as a single event with a unique trigger time. Synchronization to the VDAQ2 master clock enables seamless coincidence analyses with the cryogenic detectors.

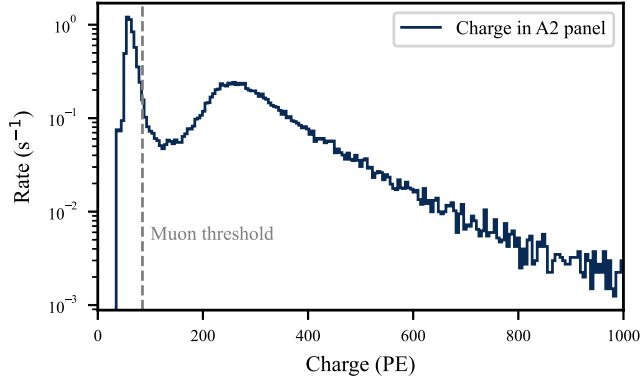

FIG. 3. Distribution of detected charge in the A2 muon veto panel (top position) taken as an example. It features two maxima, which, respectively, correspond to the low-energy gamma radiation and the Landau distribution of muon signals. The vertical dashed line indicates the chosen charge threshold used to classify an event as a muon, corresponding to approximately 3 MeV.

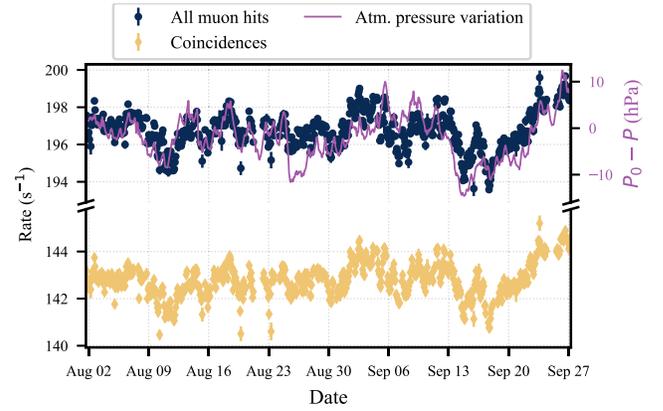

FIG. 4. Stability of the muon veto event rate over time, showing both the total muon rate and the rate of coincidences between at least two panels. The small variations are observed to be correlated with atmospheric pressure changes (shown relative to the mean pressure $P_0 = 1013$ hPa).

At the beginning of the run, the SiPMs were calibrated to convert measured charge values into photoelectron units. This calibration was performed by analyzing single photoelectron spectra obtained from dark count measurements, as described in Ref. [37]. The muon veto spectra of each panel, measured during the run, exhibit a Landau-like distribution, consistent with the energy deposition of muons traversing approximately 5 cm of plastic scintillator (see Fig. 3). The low-energy portion is mostly populated by high-energy environmental gamma rays and features a maximum close to the trigger threshold. To effectively separate the low-energy gamma background from signals produced by muons, an analysis threshold was set just above this maximum. This threshold ranged from 60 to 200 photoelectrons, depending on the panel. The variation arises from differences in scintillation light yield. The selected threshold corresponds to an energy deposition of approximately 3 MeV, obtained by converting the number of photoelectrons into energy using the light yield calibration factor. This factor was determined by comparing the position of the muon-induced Landau peak in the measured data with that in the corresponding simulation.

Events were classified as muons if at least one signal in a channel exceeded the respective analysis threshold. These muon events were subsequently used in coincidence analyses with the COV and target detectors.

Throughout the commissioning run, the total muon rate—defined as the rate at which at least one MV panel detected a muon—was monitored. The measured muon rate in the UGL was $(196.7 \pm 1.0)$ s$^{-1}$, with maximal variations of approximately 3%, which are consistent with changes in atmospheric pressure, as shown in Fig. 4.

Atmospheric pressure data were obtained with the open source *meteostat* python library [46]. The rate of coincidences between at least two panels was $(142.8 \pm 0.8)$ s$^{-1}$ and showed similar variations to the total muon rate defined previously (see Fig. 4). Additionally, the rate of muons reaching individual panels was tracked to verify the uniformity and stability of the MV system's performance, and exhibited the same stable behavior and similar variations as the total and coincidence muon rates.

Coincidences between muons tagged in the COV and the MV detectors were used to estimate the MV system's efficiency, defined as the probability that a muon reaching the cryogenic detector box is correctly identified by the MV. Above 3 MeV, the event rate in the COV is dominated by muon-induced events, with a measured rate of $(0.573 \pm 0.005)$ s$^{-1}$. Considering the fraction of these events that are identified by the MV, the measured MV efficiency is $(98.6 \pm 0.1)\%$, which remained stable throughout the whole commissioning run (see Fig. 5).

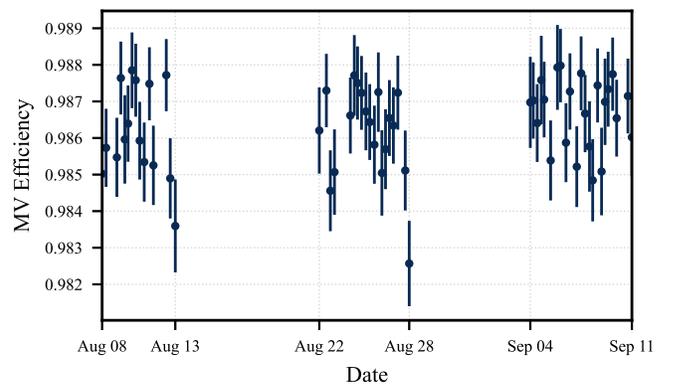

FIG. 5. Measured efficiency of the MV system over time, defined as the probability of correctly identifying a muon reaching the cryogenic detector box.

Simulations of the commissioning setup (see Sec. IV) predict a MV efficiency of 99.0%. The MV efficiency determined in this work is slightly lower than the 99.7% efficiency estimated in Ref. [37], based on purely geometrical simulations and a different definition of the efficiency (specifically as the probability that a muon reaching the lead shielding is detected by the MV), but still meets the specification of the NUCLEUS experiment.

### C. The cryogenic outer veto detector

The output signal of the COV detector was continuously sampled at 100 kHz using the VDAQ2 system. Amplified signals exhibited a fast rise time of about 300 $\mu$s and a slower decay time of approximately 2.8 ms. Differentiating the data stream sample by sample accentuates the rapid rise of physical pulses while flattening the baseline around zero, enabling the clear identification of events [47]. Triggering can then be performed on the differentiated stream by applying a threshold set above the baseline fluctuations, ensuring efficient event detection while remaining largely insensitive to pileup events. In this work, the COV is operated as a stand-alone detector, with its data acquisition triggered independently using the method described above. This approach contrasts with the intended role of the COV as a veto detector in the final experimental setup, where its readout will be correlated offline with triggers from the cryogenic target detectors, i.e., COV data will be evaluated only in coincidence with events triggered in the cryogenic detectors.

Baseline fluctuations of the differentiated stream were continuously monitored throughout the run to assess the noise level and its stability over time in the COV detector. The baseline RMS remained stable over the first week of operation (see Fig. 6, bottom), with an average value of $(1.08 \pm 0.05)$ mV over the same period. Similarly, the event rate above the trigger threshold—set to 10 mV on the differentiated stream—was tracked to monitor detector performance, showing a stable rate of $(2.13 \pm 0.01)$ s$^{-1}$ during the same period (see Fig. 6, top).

During continuous operation, the detector's performance can show slight degradation, indicated by increased baseline noise and reduced event rates. This effect can be attributed to detector neutralization: charges trapped within the bulk and on the surfaces create a counter electric field opposing the applied bias field. As a result, the effective electric field strength is reduced, leading to worse detector performance. The timescale for the neutralization process depends strongly on factors such as detector purity, surface quality, infrared environment, and the rate of incident radiation [48]. However, performance can be fully restored by *regenerating* the electric field, which involves recombining the trapped charges. This can be achieved by exposing the unbiased detector to a radioactive source or optical photons. In this commissioning run, two different scenarios were tested to provide input on the neutralization timescale. After the first week of stable data taking, the COV detector was left unbiased for the following week without data acquisition. During this period, exposure to environmental radioactivity was expected to naturally restore its performance. When data acquisition resumed, baseline RMS values and event rates were consistent with the original stable levels (see Fig. 6). In contrast, after the second COV data-taking period the detector was kept biased, even though no data were acquired, to monitor potential performance degradation and determine when regeneration would be necessary. When data taking resumed after one more week, a clear deterioration in performance was observed (see Fig. 6), indicating that regeneration was required. For long-term operation at Chooz, the COV performance will be continuously monitored, with regular regeneration with optical photons foreseen. Importantly, this regeneration procedure is relatively fast (on the order of a few hours) and, therefore, not expected to impact the experiment's duty cycle.

The energy of the pulse is reconstructed by integrating the samples above the threshold. Energy calibration was performed before the beginning of the run using a $^{137}$Cs source—emitting a gamma ray at 662 keV—positioned just below the cryostat inside the outer passive shield (see Fig. 7, top). The resolution of the $^{137}$Cs peak was measured to be approximately $(20.3 \pm 0.8)$ keV (full width half maximum). The calibration was further refined using gamma-ray lines originating from natural radioactivity present in the background spectrum (see Fig. 19). The stability of the calibration was monitored by studying the background spectrum every 8–12 hours, and the detector response remained stable at the few-percent level throughout one week of data taking.

The trigger and reconstruction efficiency was studied by adding reference pulses with the same shape as those measured and amplitudes tuned to well-defined energies into an empty stream featuring noise frequency spectrum and amplitude consistent with the measured data. The analysis of the resulting stream allowed the determination of the combined trigger and reconstruction efficiency as a function of energy, as shown in the bottom panel of Fig. 7. The inset of the same figure compares the performance in terms of mean noise RMS for the discrete differentiation method and the matched filtering approach [49,50]. Using discrete differentiation, an average noise RMS of $(3.4 \pm 0.8)$ keV was achieved. Applying a matched filter improved this by a factor of 3, reducing the noise RMS to $(1.2 \pm 0.1)$ keV. This corresponds to a potential energy threshold—defined as five times the noise RMS and corresponding to 50% trigger efficiency in the matched filter case—of about 6 keV. In contrast, for the discrete differentiation method, the same threshold definition ($5 \times$ RMS) results in a trigger efficiency close to 0%, with the 50% efficiency point only reached at around 40 keV. The trigger efficiency curves in the bottom panel of Fig. 7 clearly illustrate this difference and highlight that the improvement in the energy threshold achieved with matched filtering exceeds the factor-of-three reduction observed in the noise RMS. However, this method requires a larger time window for event triggering and energy determination, which makes it more susceptible to pileup events and results in a reduced trigger efficiency compared to the differentiation approach, as also shown in the bottom panel of Fig. 7. While the differentiation method achieved a flat, nearly 100% trigger efficiency above 60 keV, the matched filter method showed a reduced efficiency at the plateau, remaining below 95%. When the COV detectors are used as a veto, this issue can be mitigated by searching for coincident events in the COV detectors only when a trigger occurs in the cryogenic target detectors. For the final NUCLEUS setup at Chooz, both high efficiency and a very low energy threshold ($\lesssim 5$ keV) are essential for the COV [25]. However, in this study, where the focus is on using the COV detector as a stand-alone system to investigate the background environment, a low energy threshold is not a strict requirement. Consequently, the differentiation method described above is employed in the following analysis.

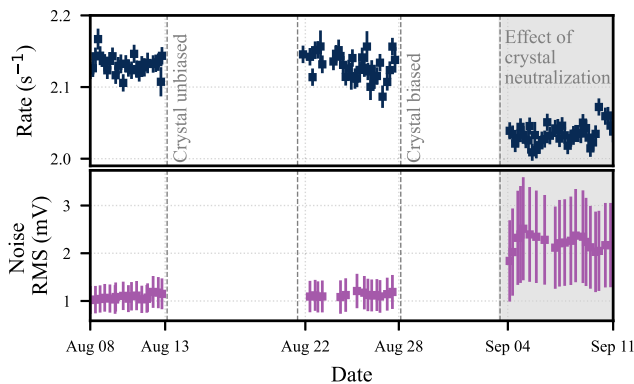

FIG. 6. Stability of the COV detector over time. The top panel shows the event rate above the trigger threshold, while the bottom panel displays the baseline RMS. Previous to the last data-taking period, indicated by the gray-shaded region, the COV crystal was not regenerated and worse performance was observed.

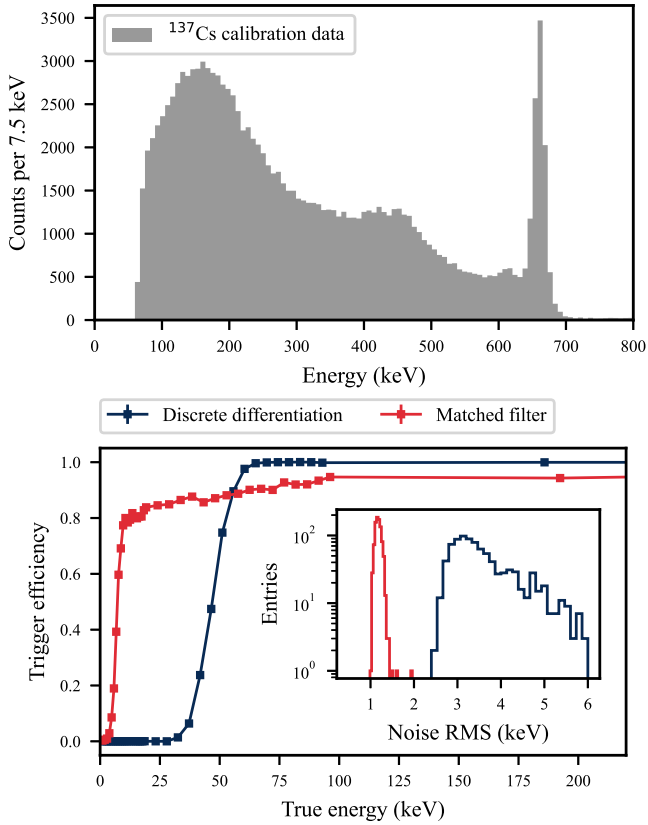

FIG. 7. Top panel: energy spectrum obtained with the COV detector during calibration with a $^{137}$Cs source positioned just below the cryostat. Triggering and energy reconstruction were done with the discrete differentiation method introduced in the text. Bottom panel: comparison of the trigger efficiency obtained using discrete differentiation (blue) and a matched filter (red) as a function of energy for the COV detector. The inset shows the corresponding noise RMS distributions.

## D. The cryogenic target detectors

### 1. Data processing and event selection

The data streams of the cryogenic target detectors were continuously sampled at 10 kHz for the CaWO$_4$ detector and 50 kHz for the Al$_2$O$_3$ detector. The higher sampling rate for Al$_2$O$_3$ was required to resolve its much faster pulse response, with time constants roughly 10 times shorter than those of CaWO$_4$. This difference arises from the use of different TES designs in the two detectors, resulting in distinct pulse shapes. A representative example is shown in Fig. 8, where a typical signal pulse is fitted using the thermal model from Ref. [51].

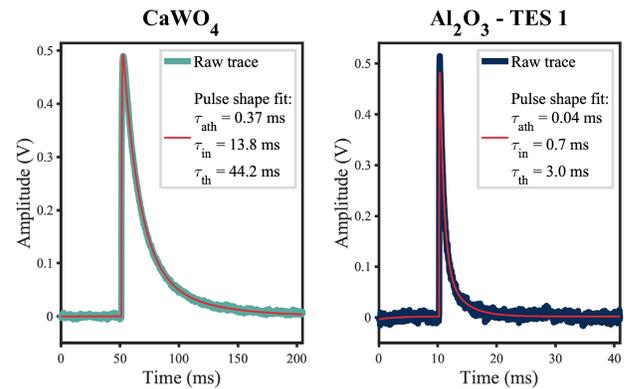

FIG. 8. Fit of the pulse shape model [51] for CaWO$_4$ (left) and Al$_2$O$_3$ (right). Indicated are the athermal phonon lifetime $\tau_{ath}$, the relaxation time of the thermometer $\tau_{in}$, and the relaxation time of the crystal $\tau_{th}$, assuming the detector is operating in calorimetric mode. The fraction of the athermal component is determined to be 80% of the total amplitude.

In addition to particle pulses, the data stream contained artificial pulses generated by injecting controlled heat onto the detector surface. These heater pulses were periodically injected—every 6 s for the $CaWO_4$ detector and every 4 s for the $Al_2O_3$ detector—at different amplitudes. They were used to monitor the detector response over time across various energies within its dynamic range (see Sec. III D 3). A hardware flag was recorded to facilitate identification during the analysis.

LED calibration data were acquired under the same conditions as the physics data, with the addition of LED-generated pulses, which were saved with a hardware flag for unambiguous identification. Besides providing energy calibration (see Sec. III D 2), LED pulses covering the full energy range from the trigger threshold to detector saturation were used to characterize the analysis and assess efficiencies (see Sec. III D 4).

The triggering of the cryogenic target detectors was performed offline using a matched filter to maximize the signal-to-noise ratio and to achieve the lowest possible energy threshold [50], which is a priority for CE$\nu$NS detection. Before processing the data, an average pulse template and an average noise power spectrum were constructed, both of which were required for the matched filter calculation.

The pulse template was obtained by averaging several hundreds of well-reconstructed signals. However, due to the low event rate in the data, a sufficient number of particle pulses was typically unavailable in a single file. To compensate for this, LED pulses were used, as their shape closely resembled that of particle pulses. In contrast, heater pulses exhibited a different pulse shape and were, therefore, unsuitable for this purpose. The validity of this approach was tested by comparing the shapes of different pulse populations—particle, LED, and heater pulses. A selected sample from each category was fitted to the thermal model, and the resulting time constants were compared. Results are summarized for the $CaWO_4$ detector as an example in Table I. The extracted time constants show that particle and LED pulses have similar rise and decay times, with $\tau_{ath} = (0.37\pm0.06)$ ms and $(0.51\pm0.02)$ ms, and $\tau_{th} = (44.2\pm3.8)$ ms and $(43.8\pm0.7)$ ms, respectively. Heater pulses exhibit a notably slower rise time [$\tau_{ath} = (1.25\pm0.01)$ ms] and a slightly shorter decay time [$\tau_{th} = (40.3\pm0.4)$ ms].

TABLE I. Time constants extracted from the fit of the thermal model to a selected sample of particle, LED, and heater pulses in the $CaWO_4$ detector. The central value is taken as the median of the fitted parameters' distribution and the uncertainty as the standard deviation of the same distribution.

|          | $\tau_{ath}$ (ms) | $\tau_{in}$ (ms) | $\tau_{th}$ (ms) |
|----------|-------------------|------------------|------------------|
| Particle | $0.37\pm0.06$     | $13.8\pm1.3$     | $44.2\pm3.8$     |
| LED      | $0.51\pm0.02$     | $13.0\pm0.1$     | $43.8\pm0.7$     |
| Heater[a] | $1.25\pm0.01$    | $15.6\pm0.1$     | $40.3\pm0.4$     |

[a]Although the underlying physics governing the shape of artificial heater pulses differs from that of particle-induced pulses, the same modeling approach and parameter labeling are adopted for consistency.

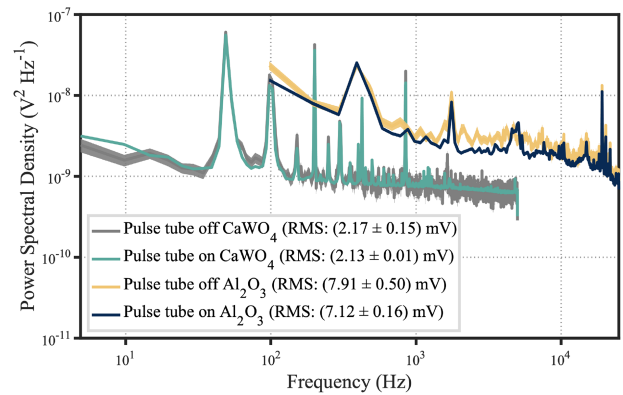

FIG. 9. Impact of the pulse tube on the noise power spectra for the $Al_2O_3$ (TES1) and $CaWO_4$ detectors.

The noise power spectrum in the frequency domain was derived by averaging the Fourier transform of baseline noise traces selected from the data stream. An example of the noise power spectrum obtained for the two detectors is shown in Fig. 9. The impact of the pulse tube on the detector noise was investigated by short measurements taken with the pulse tube switched off, and no significant effect was observed.

The triggering of events consisted of identifying all pulses in the filtered data streams whose amplitude exceeded a set threshold. This trigger threshold was typically set to five times the baseline resolution. The baseline resolution itself was estimated by applying the matched filter to a large set of noise traces and sampling the noise amplitude at a fixed position. The resulting distribution followed a Gaussian shape centered at zero, with its standard deviation providing the resolution estimate. The baseline resolution of both detectors was continuously monitored throughout the run to allow for adjustments in the analysis in response to any potential noise changes (see Sec. III D 3). Since the resolution remained stable throughout, all baseline resolution values were averaged to obtain an overall estimate: $(5.7\pm0.2)$ eV for the TES1 and $(5.5\pm0.2)$ eV for the TES2 of the $Al_2O_3$ detector and $(6.2\pm0.3)$ eV for the $CaWO_4$ detector. These values correspond to 5$\sigma$ energy thresholds of approximately 28–31 eV. The values reported here were calibrated with the LED calibration for the $Al_2O_3$ detector and the copper x-ray calibration for the $CaWO_4$ detector (see Sec. III D 2). The quoted uncertainties reflect only the statistical uncertainty from the baseline resolution determination. An additional systematic uncertainty of 25% applies to the energy calibration and should be considered in the interpretation of these baseline resolutions and energy thresholds (see Sec. III D 2).

Pulse amplitudes were determined from the maximum value of the filtered time series. Within the detector's linear response range, this amplitude is assumed to be proportional to the deposited energy. However, nonlinearities of up to 10–20% are possible even within this range, as discussed in Ref. [52]. The calibration of the matched filter amplitude is discussed in the next section.

In addition to the trigger time and pulse amplitude, several pulse shape parameters were calculated to identify valid pulses and reject artifacts. These included the pulse rise time, decay time, pretrigger baseline evaluation, baseline after pulse decay, and the $\chi^2$ value of the matched filter, which essentially quantifies pulse deformation relative to the template.

Pulse shape cuts were applied to remove artifacts and distorted pulses. One example of such artifacts is flux quantum losses (FQLs) in the SQUID [53], which can occur when rapid signal changes exceed half of a flux quantum, causing the SQUID to jump to a different working point and resulting in a baseline shift of approximately 1 V. About 340 triggers per day were associated with FQLs in the CaWO$_4$ detector, and about 430 triggers per day in the Al$_2$O$_3$ detector. When several consecutive FQL events occur, the baseline may eventually reach the limit of the amplifier, causing a SQUID reset. This results in a baseline jump of about 8 V. About 40 triggers per day were associated with SQUID reset in the CaWO$_4$ detector, and about 66 triggers per day in the Al$_2$O$_3$ detector. Together, FQLs and SQUID resets accounted for approximately 25% of the total triggers in CaWO$_4$ and 50% in Al$_2$O$_3$. Another type of rejected pulse occurs when pileup events cause a decaying voltage baseline within the recording window, leading to poor amplitude reconstruction.

Three independent analysis frameworks were used to analyze the cryogenic detector data: Diana (analysis 1), a c++ software originally developed for the CUORE/CUPID experiments [54,55] and adapted to NUCLEUS [56], Cait (analysis 2), a Python 3 package designed for processing raw data from cryogenic particle detectors [57] and ported to NUCLEUS, and CryoLab (analysis 3), a MATLAB-based software specifically developed for NUCLEUS data analysis [52]. Each detector's data was analyzed using at least two of these frameworks, yielding fully compatible results. In the following sections, results from different analyses will be presented.

### 2. Energy calibration

During this commissioning run, no radioactive calibration sources were used to not bias the background rate measurement. However, the extended duration of the detector operation allowed the observation of copper x-ray lines activated by muons passing through the setup. Copper, the most abundant material near the detectors, features a prominent K$\alpha$ x-ray line at 8.04 keV and a weaker K$\beta$ line at 8.9 keV [58].

Due to the low saturation level of the Al$_2$O$_3$ detector (~4 keV) and high flux quantum loss rate above this energy, these lines cannot be observed in its energy spectrum. In contrast, in the energy spectrum of the CaWO$_4$ detector, with a larger dynamic range (saturation ~25 keV), they are clearly visible above the flat background with a rate of $(6.07 \pm 0.53)$ counts per day, calculated summing both K$\alpha$ and K$\beta$ contributions (see Fig. 10). Since these x rays are expected to be in coincidence with muons, the energy spectrum in coincidence with the MV detectors (see Sec. III D 5) was analyzed to estimate the calibration constant. The measured positions of the copper x-ray lines yielded a calibration factor $C = (17.25 \pm 0.03)$ eV mV$^{-1}$ for the CaWO$_4$ detector, determined at an energy of approximately 8 keV and linearly extrapolated to all energies in the subsequent analysis.

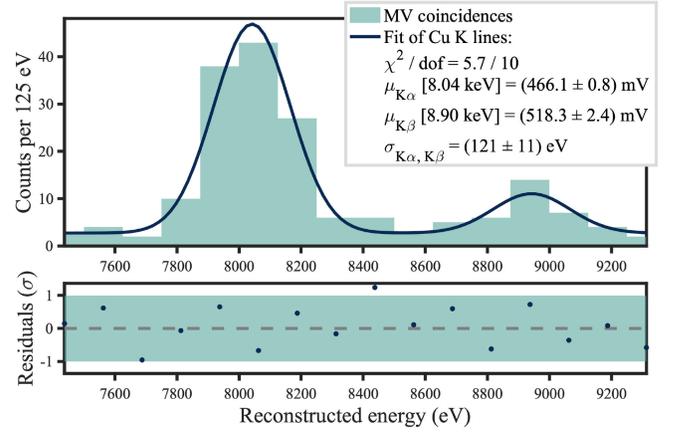

FIG. 10. Energy spectrum obtained with the full exposure of the CaWO$_4$ detector in coincidence with muons. The best fit of the copper lines, modeled as two Gaussian distributions over a flat background component, is shown. A rate of $(6.07 \pm 0.53)$ counts per day is observed, summing both K$\alpha$ and K$\beta$ contributions.

An optical calibration system was also deployed in this commissioning run. It uses a controlled LED to generate photon signals of varying intensity. This nonintrusive method enables energy calibration over a broad range by analyzing photon statistics and relative pulse amplitudes [32,59]. As previously mentioned, LED calibrations were performed at the beginning and end of weekly data-taking periods, ensuring continuous monitoring of the detector response.

In order to perform the energy calibration, the different LED-induced signals were identified in the histogram of matched filter amplitudes. Assuming a Gaussian distribution, their mean $\mu_\gamma$ and variance $\sigma^2$ were evaluated. Then, $\sigma^2$ was fitted as a function of $\mu_\gamma$ with the following relation:

$$\sigma^2 = \frac{1}{C} \cdot \epsilon_\gamma \cdot \mu_\gamma + \sigma_0^2, \qquad (1)$$

where $C$ is the calibration factor, $\epsilon_\gamma$ is the energy of the single photon (here 4.86 eV), which is a fixed value determined by the LED wavelength, and $\sigma_0$ is the detector intrinsic noise contribution to the observed resolution. Both parameters were left free in the fit. The fitted values of the detector resolution are compatible with those evaluated from the distributions of the noise amplitude.

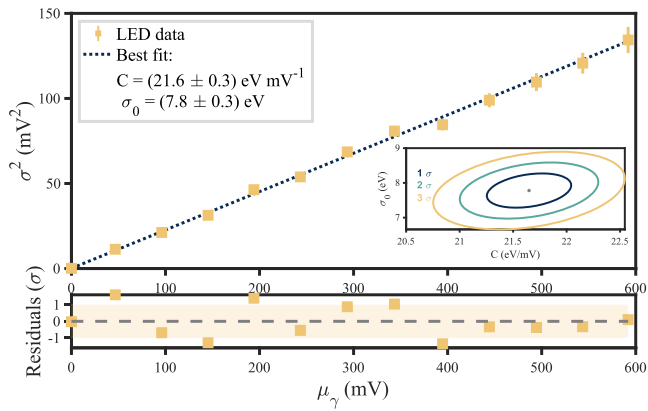

FIG. 11. Fit of Eq. (1) to the LED data acquired on September 10 with the CaWO$_4$ detector. Each data point's statistical uncertainty was added to an uncorrelated uncertainty to normalize the $\chi^2$ per degree of freedom to unity. This procedure accounts for possible systematic uncertainties in the extraction of the calibration parameters from the fit. The inset shows the correlation between the two fit parameters.

An example of an optical calibration for the CaWO$_4$ detector is shown in Fig. 11. Since no significant detector response variations were observed throughout the run (see Sec. III D 3), all LED calibrations were averaged, resulting in a mean calibration factor of $(21.5 \pm 0.2)$ eV mV$^{-1}$ for the CaWO$_4$ detector. For the Al$_2$O$_3$ detector, only the optical calibration was available, and each week's data were calibrated using the corresponding LED measurement taken at the start of that week.

Potential systematic uncertainties in the optical calibration procedure were studied for both detectors, by testing the effects from phonon statistics and detector nonlinearities [60] on the energy calibration factor. No significant deviation was observed, indicating that these effects do not introduce a substantial systematic uncertainty.

For the CaWO$_4$ detector, the calibration constant obtained from the copper x-ray lines is $(17.25 \pm 0.03)$ eV mV$^{-1}$, while the LED calibration yields $(21.5 \pm 0.2)$ eV mV$^{-1}$. This corresponds to a discrepancy of approximately 25% with a statistical significance of about 21$\sigma$. The origin of this discrepancy is currently unknown and constitutes the dominant systematic uncertainty in the energy calibration. To investigate the observed discrepancy, the NUCLEUS Collaboration is conducting a systematic study of calibration methods at very low energies. In addition to the optical calibration, an x-ray fluorescence source has been developed to extend the calibration range down to sub-keV energies, providing discrete lines between 600 eV and 6 keV. In [52], a significant detector nonlinearity was observed using this technique, which was, however, neglected in this work.

An x-ray fluorescence source could also allow for cross-validation and benchmarking of the optical calibration, although such a comparison has not yet been performed. Furthermore, within the CRAB project, an alternative calibration approach is being explored using sub-keV nuclear recoils induced by MeV gamma rays following thermal neutron capture [61,62]. This method provides a calibration using sub-keV nuclear recoils distributed throughout the detector volume, closely mimicking the signal from neutrinos. In contrast, x-ray fluorescence produces mostly surface electron recoils at higher energies, and LED pulses, while volumetric, probe electron recoils rather than nuclear recoils. Anchoring the regular LED calibration to a CRAB nuclear-recoil measurement will allow reduction of systematic uncertainties associated with differences between electron and nuclear recoil responses in the energy region of interest.

### 3. Detector stability

A thorough evaluation of the cryogenic target detectors' stability was performed, focusing on several key observables. Anticipating the results, all indicators confirmed stable detector operation throughout the entire commissioning run, with no significant deviations observed. Consequently, no data were excluded from the analysis based on stability criteria.

The overall event rate of triggered events—calculated above a threshold of 1 keV in order to exclude the LEE region—remained stable throughout the run in both detectors, as illustrated in Fig. 12.

To further assess the long-term stability, the amplitudes of the eight different heater pulses periodically injected during data taking were monitored. Figure 13 shows the reconstructed amplitudes of the four lowest amplitude heater pulses as a function of time, in the CaWO$_4$ detector taken as an example. A stability better than 1% was observed in both detectors throughout the run, indicating a stable detector response over the entire data-taking period.

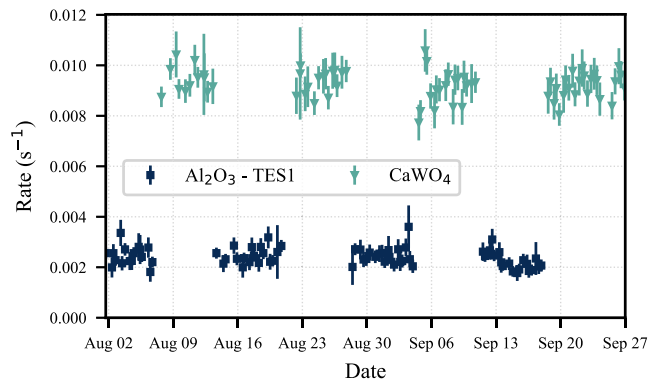

FIG. 12. Stability of the event rate (above 1 keV) recorded by the target detectors over the commissioning period.

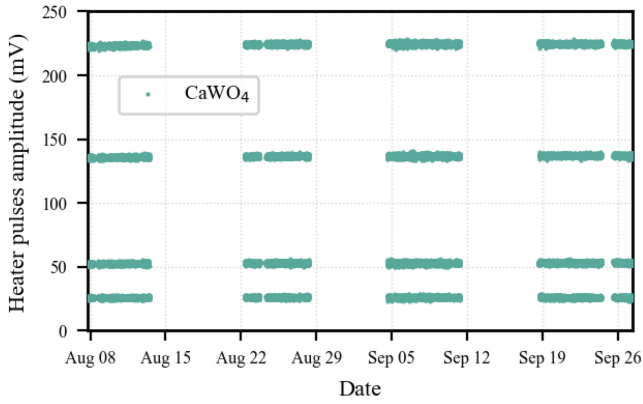

FIG. 13. Stability of the reconstructed amplitudes of the four heater pulses with the lowest amplitudes over the commissioning run in the CaWO$_4$ detector.

Additional confirmation of stable working conditions comes from the multiple LED optical calibrations performed throughout the run. Figure 14 shows the extracted calibration constants from independent LED calibrations, all of which are consistent within uncertainties. This agreement further verifies that the TES working points remained unchanged between different data-taking periods.

The noise level of the detectors—quantified in terms of baseline resolution—was also continuously monitored. As shown in Fig. 15, both detectors exhibited stable noise behavior, with only a slight increase observed toward the end of the run—approximately 10% in one TES of the Al$_2$O$_3$ detector and 12% in the CaWO$_4$ detector. This increase was not considered critical, and no data were excluded from the analysis. However, larger noise fluctuations can lead to a higher rate of low-energy noise triggers near the detection threshold, effectively increasing the threshold itself. A discussion of how this is handled in the analysis is beyond the scope of this work, as the focus remains on the energy region well above the threshold.

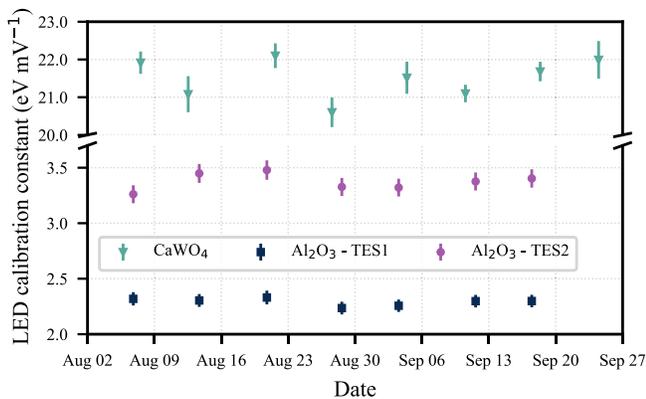

FIG. 14. Stability of the LED calibration constant for the CaWO$_4$ and Al$_2$O$_3$ detectors throughout the commissioning run, proving a stable detector response over time.

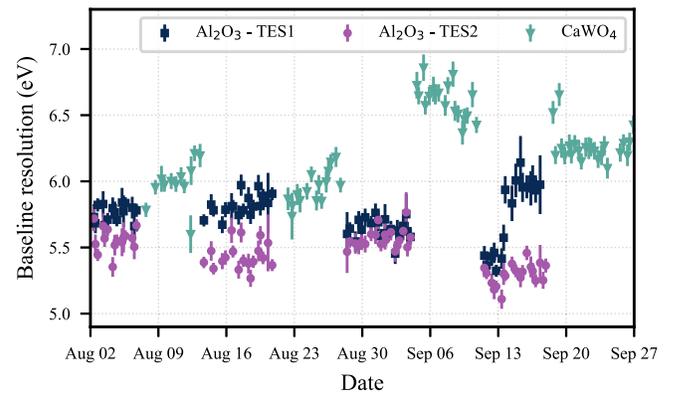

FIG. 15. Baseline noise resolution of the CaWO$_4$ and Al$_2$O$_3$ detectors over time. Both detectors exhibit stable noise behavior, with nonrelevant variations toward the end of the run.

## 4. Data analysis efficiency

The efficiency of the data analysis is influenced by three main factors: triggering, reconstruction, and cleaning. In this work, two distinct approaches were used to evaluate the efficiency: one based on LED data and another based on simulations.

LED pulses offer a unique opportunity for efficiency assessment, as the characteristics of each event, such as energy and time, are precisely known. By processing the LED data using the same analysis procedures applied to physics data, the number of triggered events and their reconstructed energies can be compared with the expected values, providing a direct estimate of the trigger and reconstruction efficiency as a function of energy. Furthermore, by applying pulse shape cuts to the LED data, the cleaning efficiency can be evaluated by comparing the number of LED events passing the pulse shape cuts to the number of events that were initially triggered.

Figure 16 shows the efficiencies as a function of the energy for the CaWO$_4$ detector, which is taken as an example. The trigger efficiency at low energy is consistent with the expected effect of Gaussian resolution. The efficiency as a function of the average deposited energy ($E$) can be fitted with the following function:

$$\epsilon_{\text{trig}}(E) = \frac{A}{2} \cdot \left[ \text{erf}\left( \frac{E - E_0}{\sqrt{2 \cdot (\sigma_0^2 + E \cdot \epsilon_\gamma)}} \right) + 1 \right], \quad (2)$$

where $A$ models the energy-independent efficiency loss, $E_0$ is the energy at which 50% of $A$ is reached, and the denominator contains the resolution. Since LED pulses are used, the detector baseline resolution $\sigma_0$ is corrected with an additional contribution from Poisson statistics, proportional to the deposited energy, $E \cdot \epsilon_\gamma$.

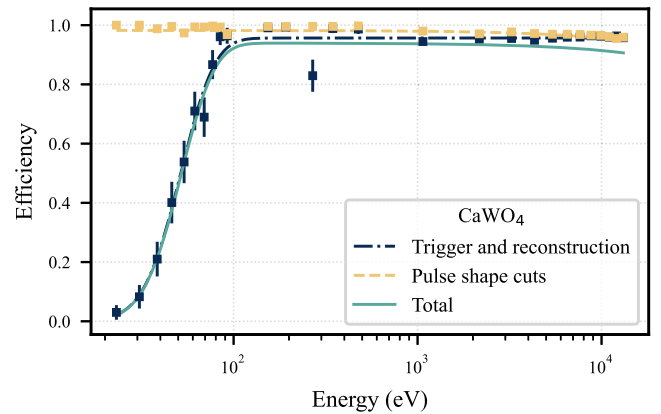

FIG. 16. Efficiencies of trigger, reconstruction, and pulse shape cuts for the CaWO$_4$ detector, as obtained from the analysis of LED pulses (see the text for more). The uncertainty bars are the $1\sigma$ statistical fluctuations given by a binomial distribution.

For the CaWO$_4$ detector, the trigger efficiency was estimated as $A = (95.66 \pm 0.19)\%$, and for the Al$_2$O$_3$ detector, as $A = (99.54 \pm 0.11)\%$. The efficiency does not reach 100% due to the presence of heater pulses in the data stream, which introduce a limited dead time in the trigger algorithm. The higher efficiency observed in the Al$_2$O$_3$ detector is attributed to its faster pulse response, which reduces the duration of this dead time (see Fig. 8).

The efficiency of the pulse shape cuts was estimated in a similar manner. For the CaWO$_4$ detector, the efficiency as a function of energy was fitted with a linear function: $\epsilon_{\text{cuts}}(E) = C + B \cdot E$, to account for the energy dependence of some pulse shape cuts. A maximum efficiency of $C = (98.24 \pm 0.12)\%$ and a linear decrease of $B = (-0.27 \pm 0.02)\%$ keV$^{-1}$ were obtained. This results in a maximum variation of about 3% in efficiency across the full energy range—from the energy threshold to 13 keV. For the Al$_2$O$_3$ detector, an almost constant efficiency of above 99% was measured.

A second method for evaluating efficiency was used as a cross-check with compatible results. This involves stream simulations, where reference pulses with well-defined energies are added to an empty data stream containing noise levels consistent with the actual measurements. Since the characteristics of the simulated pulses—such as energy and timing—are known, the efficiency can be estimated in a similar manner to the LED data analysis. However, a limitation of the simulation method is that it cannot fully replicate certain detector effects, such as nonlinearities or position-dependent variations. As a result, the efficiency measured through simulations may be slightly overestimated and it is not useful when LED data are available.

### 5. Veto coincidence definition and time resolution

A time coincidence between a cryogenic target detector and one of the veto detectors was defined by requiring that a veto event falls within a narrow time window around the trigger time of the TES signal. The size of this coincidence window was determined by analyzing the distribution of time differences between events in the target detector and the closest event in the veto.

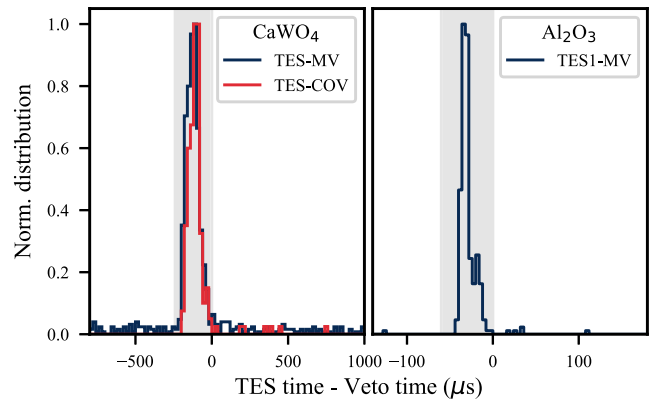

FIG. 17. Distribution of the time difference between a TES event and the nearest MV event (in blue) and COV event (in red), shown for the CaWO$_4$ detector (left panel) and Al$_2$O$_3$ detector (right panel). The gray-shaded areas indicate the time coincidence windows used in the analysis.

Figure 17 shows the time difference distribution for the CaWO$_4$ detector. A mean negative offset of approximately 120 $\mu$s and a standard deviation of about 40 $\mu$s were observed for both the MV and COV vetoes. The negative sign of the offset is to be attributed to instrumental and not physical effects. The fact that the same standard deviation is observed for MV and COV coincidences suggests that it is primarily limited by the resolution of the TES signal time reconstruction. Based on these observations, the time coincidence window was defined from $-240$ $\mu$s to 0 $\mu$s, capturing approximately 99.7% of true TES-veto coincidences. The level of random coincidences within the defined time window was estimated using the measured muon rate, yielding expected accidental coincidences of $\approx 197$ Hz $\cdot$ 240 $\mu$s = 4.7%, assuming the muon rate is dominant compared to the cryogenic detector rate. In contrast, random coincidences with the COV were found to be negligible.

For the faster Al$_2$O$_3$ detector, a standard deviation of the time difference distribution of approximately 7 $\mu$s was measured, along with a negative offset of 30 $\mu$s. Based on this, a narrower coincidence window from $-60$ $\mu$s to 0 $\mu$s was chosen, capturing more than 99.9% of true TES-veto coincidences. The estimated accidental coincidences with the MV are approximately 1.2%.

The matched filter enables a more precise estimation of the TES signal time than the raw data. Theoretically, the timing precision of the matched filter is expected to scale as $\propto E^{-1}$, where $E$ is the signal energy [63]. This expected scaling behavior was checked with LED calibration data. Figure 18 shows the reconstructed time precision of LED pulses as a function of their energy for the Al$_2$O$_3$ detector taken as an example. The relation was fitted with a general function

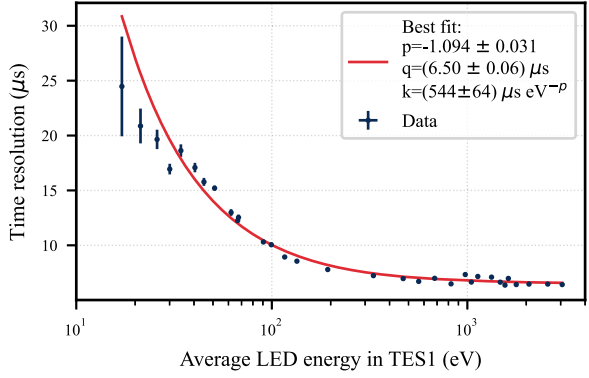

FIG. 18. Time resolution as a function of the average reconstructed energy evaluated using LED pulses for TES1 of the $Al_2O_3$ detector. A good agreement with the expected $\propto E^{-1}$ scaling is observed.

$$\sigma_t(E) = k \cdot E^p + q, \quad (3)$$

where $E$ is the energy, $k$ is a normalization constant, $q$ is a constant offset, and $p$ represents the expected power-law scaling of the time resolution. The best-fit result, $p = -1.09 \pm 0.03$, is in agreement with the theoretical expectation of $p = -1$ within $3\sigma$. Furthermore, the constant offset obtained from the fit is consistent with the standard deviation of the time difference distribution observed for physical coincidences with the MV, confirming that the dominant limitation arises from the time resolution of the target detectors. This study confirms that the chosen coincidence windows remain valid even when considering the energy dependence of the time resolution down to the analysis threshold.

Based on the time coincidence windows established in this work and the expected muon rate at the VNS of 325 Hz, the resulting muon-induced dead time at the Chooz site is estimated to be between 2% and 8%, depending on the detector response.

## IV. BACKGROUND MEASUREMENTS AND COMPARISON TO SIMULATIONS

The expected particle background contributions in the target detectors used in this work were evaluated through extensive simulations. For this purpose, the NUCLEUS simulation framework—originally developed to estimate background contributions for the NUCLEUS experiment at the Chooz site and described in Refs. [25,30]—was updated for the commissioning setup at TUM. Specifically, the simulation geometry was adjusted to reflect the differences between the NUCLEUS experiment at Chooz and the experimental setup used in this work. Additionally, the UGL site was incorporated into the simulation framework. The simulation parameters were tuned and validated using data collected from all detectors (MV, COV, and the target detectors $CaWO_4$ and $Al_2O_3$).

In addition to the data presented in this work, an important input for tuning the simulations was obtained from a short follow-up run with reduced active and passive shielding. In this run, the left and right shielding carts were each moved approximately 1 m away from the cryostat position along the rail supports. This configuration significantly increased the contribution of ambient gamma radiation—with simulations predicting an increased rate of ambient gamma-ray background events in the COV detector by a factor of approximately 30—while rendering the impact of radioactive contamination negligible, providing a valuable benchmark for modeling different background sources. The simulations accounted for atmospheric muons, ambient gamma radiation, and radioactive contamination in the setup materials. The latter was measured at the STELLA (SubTERranean Low-Level Assay) screening facility at the Laboratorio Nazionale del Gran Sasso of INFN [64].

A key distinction between the VNS and UGL sites is the amount of overburden, with VNS located at approximately 3 m w.e. and UGL at 10 m w.e., as determined in this work from measured muon rates in the MV detectors. This affects the cosmic-ray-induced background, as atmospheric muons are attenuated by a factor of approximately 2.3 at the UGL, and the contribution from atmospheric neutrons becomes negligible. The overburden at the UGL was incorporated into the atmospheric muon simulations (see Ref. [25] for details on the atmospheric muon generator). The average soil density was adjusted by comparing simulated and measured counting rates from the 29 MV detectors. The best agreement (within a few percent in most cases) was achieved with an assumed average soil density of $1.95 \text{ g cm}^{-3}$. This density also resulted in an atmospheric muon prediction that closely matched the energy spectra in the COV when measured in coincidence with the MV, for both shielding configurations, further validating the simulation (see Fig. 19).

The ambient gamma-ray background was characterized using measurements taken in the reduced-shielding configuration. Subtracting the atmospheric muon contribution from the COV data yielded a residual spectrum at $E < 3$ MeV, entirely dominated by ambient gamma radiation. This spectrum was used to determine the fluxes of the three main ambient gamma components: $^{238}U$, $^{232}Th$, and $^{40}K$ (see Ref. [25] for details). The resulting model was then applied to predict the background in the COV detector under full-shielding conditions. In this case, simulations also included contributions from radioactive contamination in the setup materials, which were negligible in the reduced-shielding configuration. The combined model—accounting for atmospheric muons, ambient gamma radiation, and material contamination—showed excellent agreement with COV data without the need to rescale any of the simulated components (see Fig. 19), with deviations below 20% across the entire energy range from 60 keV to 15 MeV. After subtracting the muon and gamma components from the COV spectrum, the residual background is compatible with expectations from radioactive contamination in the materials, further validating the input parameters from the screening measurements. All these results serve as a strong validation of the simulations. Finally, the same full model was used to predict the background in the $CaWO_4$ and $Al_2O_3$ target detectors.

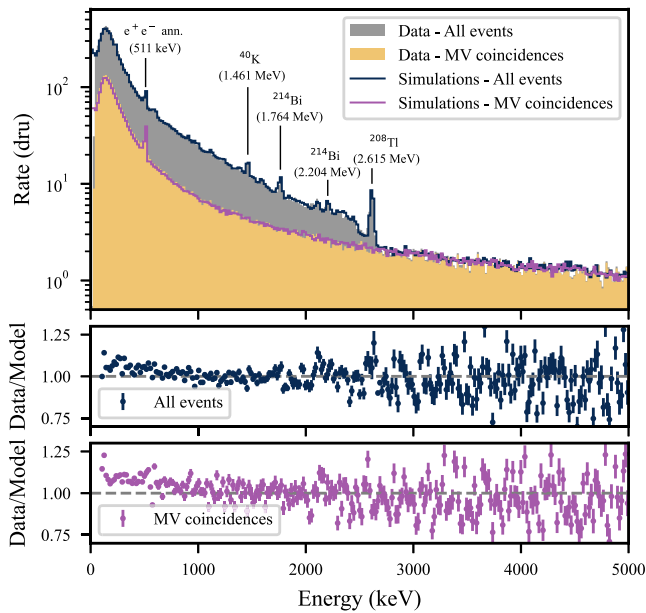

FIG. 19. Energy spectrum recorded with the COV detector compared to simulations accounting for atmospheric muons, ambient gamma radiation, and material contamination (see the text for details). In the top panel, the event rate as a function of energy is shown for data and simulations, both for all events and for events in coincidence with the MV. The most prominent background gamma lines are labeled. The two lower panels display the ratio of data to simulation, for all events in blue and for events in coincidence with the MV in purple.

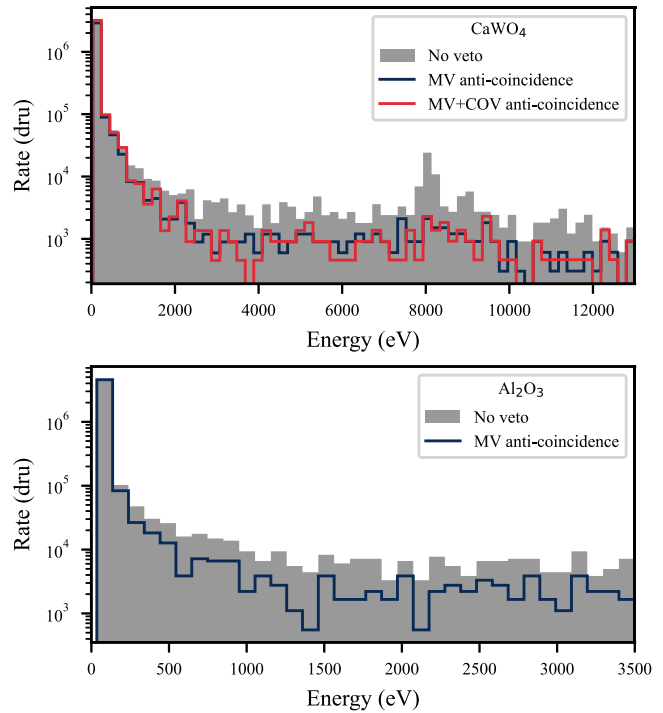

FIG. 20. Energy spectra recorded with the $CaWO_4$ detector (top) and with the $Al_2O_3$ detector (bottom), before and after the anticoincidence cut with the vetoes. The rate is corrected for the analysis efficiency.

Figure 20 presents the energy spectra measured in the two cryogenic target detectors, following the analysis procedure presented in the previous section. They are expressed in differential rate units (dru), which correspond to counts per keV kg day.

The measured spectra appear largely flat across the keV energy range, with the notable exception of the characteristic copper fluorescence lines observed in the $CaWO_4$ detector. After application of the anticoincidence cut with the veto detectors, the spectra in the keV range remain flat, and the copper lines in the $CaWO_4$ detector are fully suppressed, consistent with their expected origin from atmospheric muon interactions. Both detectors exhibit a sharp increase in the event rate at low energies—below 2 keV for $CaWO_4$ and below 1 keV for $Al_2O_3$. A similar sharp increase in the event rate at very low energies—commonly referred to as the LEE—has been reported in various low-threshold experiments and its origin remains unclear [26]. A dedicated study focusing on the LEE, using both these commissioning data and additional past measurements, will be the subject of a forthcoming publication. This work excludes the energy region affected by the LEE and focuses on the comparison between data and simulations at higher energies—above 1 keV for $Al_2O_3$ and above 2 keV for $CaWO_4$.

A comparison between measured data and background simulations is presented in Fig. 21. For both detectors, the simulated spectrum is dominated by atmospheric muon-induced background ($\gtrsim 70\%$), while the contributions from ambient gamma radiation and material contamination are subdominant and roughly comparable.

For the $CaWO_4$ detector, background simulations predict a flat spectrum in the keV region, including the two copper lines and a mild rise in the event rate below 1 keV. In the energy interval between 2.5 keV and 7.7 keV, data and simulations agree within a factor of approximately 1.5. However, in the energy range covering the copper lines (7.7–9.5 keV), the discrepancy increases to a factor of about 2.3. Above the copper lines (9.5–13 keV), the agreement improves again, reaching a factor close to 1. The sharp increase in the event rate observed in data below 2 keV is significantly larger than the mild excess predicted by simulations.

For the $Al_2O_3$ detector, background simulations predict a nearly flat spectrum down to the threshold, along with the presence of the two copper lines. However, these lines are not accessible in the measurement due to the limited dynamic range of the detector. In the 1.4–3.5 keV region, data and simulations agree within a factor of approximately 1.5, consistent with the level of agreement observed in the flat region of the CaWO$_4$ spectrum below the copper lines. Similar to CaWO$_4$, the steep rise in the event rate below 1 keV observed in Al$_2$O$_3$ data is not accounted for by simulations and exceeds predictions by orders of magnitude.

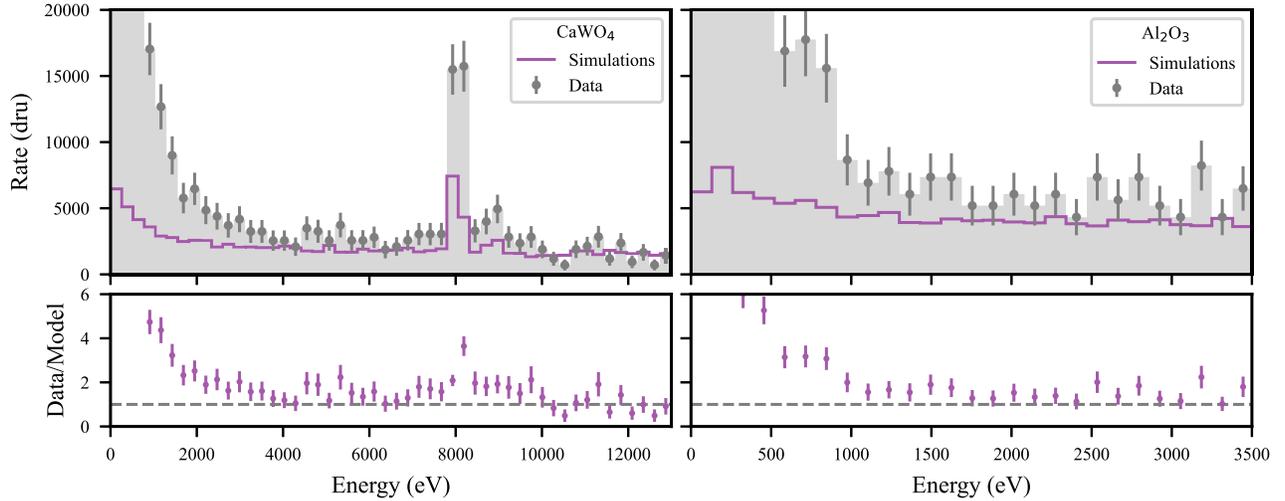

FIG. 21. Comparison between measured energy spectra and background simulations for the CaWO$_4$ (left) and Al$_2$O$_3$ (right) detectors. In the top panels, the event rate as a function of energy is shown along with the corresponding simulation. The lower panels display the ratio of data to simulation.

A quantitative summary of the event rates across different energy intervals is provided in Table II. The levels of agreement and disagreement between measured data and background simulations—described above for the data without any veto cut—remain essentially unchanged after applying the coincidence or anticoincidence cuts with the veto detectors. A broader discussion of the data simulation comparison is presented in the next section.

TABLE II. Measured event rates and corresponding background simulation predictions, both expressed in dru, along with the ratio of data to simulation. For the Al$_2$O$_3$ detector, the rate is evaluated in the flat region of the spectrum between 1.4 and 3.5 keV. For the CaWO$_4$ detector, three energy intervals are considered: 2.5–7.7 keV (below the copper lines), 7.7–9.5 keV (covering the copper lines), and 9.5–13 keV (above the lines). For each energy range, rates are shown under different veto conditions: without veto cuts, in anticoincidence and coincidence with the MV, and in combination with the outer veto (COV + MV).

| Detector | Energy range | Cut | Measured rate (dru) | Simulated rate (dru) | Data/Model |
|---|---|---|---|---|---|
| Al$_2$O$_3$ | 1.4–3.5 keV | No veto | $5979 \pm 402$ | $3964 \pm 68$ | $1.5 \pm 0.1$ |
| | | MV anticoincidence | $2300 \pm 250$ | $1459 \pm 50$ | $1.6 \pm 0.2$ |
| | | MV coincidence | $3680 \pm 316$ | $2505 \pm 47$ | $1.5 \pm 0.1$ |
| CaWO$_4$ | 2.5–7.7 keV | No veto | $2853 \pm 182$ | $1909 \pm 28$ | $1.5 \pm 0.1$ |
| | | MV anticoincidence | $967 \pm 106$ | $574 \pm 18$ | $1.7 \pm 0.2$ |
| | | MV coincidence | $1887 \pm 148$ | $1335 \pm 21$ | $1.4 \pm 0.1$ |
| | | COV + MV anticoincidence | $868 \pm 124$ | $527 \pm 17$ | $1.6 \pm 0.2$ |
| | | COV + MV coincidence | $744 \pm 115$ | $477 \pm 13$ | $1.6 \pm 0.2$ |
| CaWO$_4$ | 7.7–9.5 keV | No veto | $7223 \pm 493$ | $3084 \pm 59$ | $2.3 \pm 0.2$ |
| | | MV anticoincidence | $1411 \pm 218$ | $733 \pm 35$ | $1.9 \pm 0.3$ |
| | | MV coincidence | $5812 \pm 442$ | $2351 \pm 48$ | $2.5 \pm 0.2$ |
| | | COV + MV anticoincidence | $1431 \pm 270$ | $682 \pm 33$ | $2.1 \pm 0.4$ |
| | | COV + MV coincidence | $2351 \pm 347$ | $872 \pm 29$ | $2.7 \pm 0.4$ |
| CaWO$_4$ | 9.5–13 keV | No veto | $1650 \pm 170$ | $1571 \pm 33$ | $1.1 \pm 0.1$ |
| | | MV anticoincidence | $509 \pm 95$ | $531 \pm 24$ | $1.0 \pm 0.2$ |
| | | MV coincidence | $1141 \pm 141$ | $1040 \pm 23$ | $1.1 \pm 0.1$ |
| | | COV + MV anticoincidence | $481 \pm 113$ | $495 \pm 23$ | $1.0 \pm 0.2$ |
| | | COV + MV coincidence | $267 \pm 84$ | $374 \pm 14$ | $0.7 \pm 0.2$ |

## V. TOWARD CEνNS DETECTION: CHALLENGES AND IMPROVEMENTS

The commissioning phase at TUM aimed to validate the operational stability of the experiment and provide an initial measurement of the background using a simplified version of the NUCLEUS experiment. These objectives were successfully achieved, leading to the results presented in this work. In this section, we critically assess the implications of these findings for the CEνNS detection phase of the NUCLEUS experiment, addressing current challenges and planned improvements.

While the TUM commissioning was a crucial step, the experimental setup differed in key aspects from the final NUCLEUS configuration at Chooz. Only two cryogenic target detectors were operated ($1 \text{ CaWO}_4 + 1 \text{ Al}_2\text{O}_3$), resulting in a total active mass insufficient for CEνNS detection. Under baseline assumptions—an average anti-neutrino flux of $1.72 \times 10^{12} \text{ cm}^{-2} \text{ s}^{-1}$ and a [20, 100] eV region of interest—this configuration would yield an expected CEνNS event rate of approximately 0.015 counts per day, compared to the 0.14 counts per day for the nominal NUCLEUS configuration of $9 \text{ CaWO}_4 + 9 \text{ Al}_2\text{O}_3$ detectors. In addition, the two detectors used in this work were housed in individual copper holders, unlike the NUCLEUS configuration at Chooz, which will use silicon holders equipped with heat channel readouts. This constitutes the last piece of the NUCLEUS shielding strategy, called inner veto, designed to ultimately reject surface events and holder-related events. The COV was only partially implemented in this work, with one of the six planned high-purity germanium crystals operational, limiting its background suppression capability. Additionally, the last layer of passive shielding inside the cryostat—the $B_4C$ neutron absorber layer—was not deployed. While its absence had little impact on background suppression in this measurement—given the attenuation of atmospheric neutrons by the shallow overburden—it will play a crucial role in mitigating neutron-induced backgrounds at Chooz [25].

To build upon the progress made during commissioning, several key upgrades are planned. All six high-purity germanium crystals are currently undergoing thorough characterization at TUM, and the final COV configuration is being validated. In parallel, a minimal detector module consisting of four $CaWO_4$ detectors equipped with double-TES readout has been developed at TUM. The double readout technology, pioneered by the CRESST [65] and TESSERACT [66] experiments, has proven effective in identifying and discriminating one component of the LEE. This component consists of events that primarily couple to the sensor (TES and phonon collector), as opposed to particle events that couple to the detector phonon system. This development represents a significant step forward in enhancing the sensitivity to reactor CEνNS signals with cryogenic detectors and constitutes a natural evolution of the original NUCLEUS experimental proposal [27].

The minimal detector module will soon be tested within the complete COV shield, and both components will be integrated into the experiment during the upcoming technical run at Chooz. At this stage, the missing $B_4C$ neutron absorber layer will also be added, marking a crucial milestone in demonstrating the background levels achievable with the full NUCLEUS shielding configuration.

A low detection threshold is a key requirement for CEνNS sensitivity. In this work, energy resolutions of about 5.5 eV and 6.5 eV have been achieved with the $Al_2O_3$ and $CaWO_4$ detectors, respectively, values that meet the experiment's specifications, although significant systematic uncertainties in the energy calibration remain to be understood. Nonetheless, there are promising indications that further improvements are possible. For instance, optimizing the TES operating point on the two detectors used in this work, aimed at maximizing the signal-to-noise ratio, suggests potential improvements in baseline resolution of up to 30%. In parallel, the application of advanced matched filter techniques to the double-TES $Al_2O_3$ detector—accounting for noise correlations between sensors [67]—has demonstrated additional gains in resolution [68]. This last improvement could be extended to the $CaWO_4$ detector by adapting the double-TES design to the $CaWO_4$ crystals, a development currently underway, as mentioned above.

The detailed comparison between data and simulations presented in this work reveals a generally coherent picture across various detectors, shielding configurations, and coincidence conditions. In particular, a very good agreement was observed for the COV detector, both in the energy region dominated by ambient gamma radiation (60 keV to 2.7 MeV) and the energy region dominated by atmospheric muons (2.7–15 MeV), and for the MV detectors. These findings indicate that the simulation accurately models the atmospheric muon flux, ambient gamma background, and the complex geometry of the experimental shielding. The consistency between COV and MV data provides a strong cross-check of the atmospheric muon background, while the agreement between COV data under different shielding configurations confirms the robustness of the ambient gamma model. Across all tested configurations, including different coincidence cuts and shielding setups, the simulations correctly capture the dominant background contributions in the energy range above 60 keV. However, discrepancies persist at lower energies in the target detectors, where simulations tend to underesti the flat background component and the copper x-ray lines. This is a known challenge in low-energy rare-event searches (see, for example, [69,70]) and is likely related to limitations of Geant4 at low energies. A similar disagreement between simulations and measured data for low-energy x rays has also been reported by the CONUS experiment [71]. Initially, a larger discrepancy was observed between measured data and background simulations for the target detectors, with simulated counting rates up to a factor of 2 to 15 lower than those measured, depending on the energy region and coincidence conditions. Significant improvements were achieved by adjusting Geant4 simulation parameters, in particular by reducing particle production range cuts (to values smaller than the default) and by enabling the particle-induced x-ray emission (PIXE) process. Reducing the particle production range cut led to an increase—by up to a factor of 3—in the continuum background level and produced the low-energy spectral rise observed in data. Enabling PIXE resulted in a marked enhancement of the copper fluorescence lines in the CaWO$_4$ spectrum. While these changes significantly improved the agreement with data, they also led to a substantial increase in computing time. A practical compromise was reached by setting a stopping range of 20 nm and applying both the short-range cuts and the PIXE process only to the materials immediately surrounding the target detectors. Work is ongoing to further improve the agreement in this low-energy regime, including a systematic reevaluation of Geant4 physics lists and physics parameters and the possible integration of complementary simulation tools. While the background in the present setup is dominated by gamma rays and muons, the expected dominant background at Chooz will be due to atmospheric neutrons, which could not be probed in the current configuration. Dedicated measurements during the upcoming technical run will, therefore, be essential to validate the neutron background model.

Despite this overall agreement between data and simulations, it is important to note that in the region of interest for CE$\nu$NS detection ($\lesssim$100 eV), the background is no longer dominated by particle-induced events, but by the so-called LEE. This background component currently sets the main sensitivity limit of the experiment in the low-energy regime. The LEE dominates the trigger event rate at low energy and significantly exceeds the expected CE$\nu$NS signal at the VNS. For instance, the LEE-induced trigger rate measured in this work in the CaWO$_4$ detector is approximately 240 counts per day between 40 and 100 eV, compared to only 0.008 counts per day expected for CE$\nu$NS events in the same energy range in one CaWO$_4$ detector of the nominal NUCLEUS configuration at Chooz. This highlights the importance of developing a robust understanding and mitigation strategy for the LEE. Dedicated studies aimed at understanding the origin of the LEE and developing an effective mitigation strategy have been conducted using data from this commissioning run, along with additional measurements from the NUCLEUS Collaboration. The results of these investigations will be presented in an upcoming publication. Preliminary findings suggest that a combination of double-TES readout for all target detectors and an instrumented holder presents a promising path forward. The double-TES readout has already been shown to help discriminate TES-related events [65,66,72] and will be implemented in the minimal detector module for the upcoming technical run at Chooz. The instrumented holder—inner veto—will help discriminate the remaining LEE events, which are thought to be due to external mechanical stress on the detector or external radiation. While promising results have been obtained with the NUCLEUS prototype detector [42], the integration of this approach into the full NUCLEUS setup remains to be validated. Demonstrating the operability of NUCLEUS detectors with instrumented holders will be a key focus of the upcoming R&D phase, running in parallel with the technical run at Chooz.

## VI. CONCLUSIONS AND OUTLOOK

The commissioning of the NUCLEUS experiment at TUM has successfully demonstrated the operational stability and performance of a simplified version of the experimental setup. This phase was crucial in validating the key components and subsystems, including the cryogenic target detectors, shielding systems, and veto detectors, in preparation for the upcoming deployment at the Chooz nuclear power plant.

During the commissioning run, the experimental setup exhibited robust performance and stability across all detector subsystems. The MV system effectively identified and tagged muon events with high efficiency, while the COV provided valuable insights into the background environment near the detectors. At the same time, the two target detectors achieved energy resolutions close to the design goals and operated stably, without significant influence from vibrations of the dry cryostat. The comparison between COV data and simulations in the higher energy range (above 60 keV) confirms that atmospheric muons, ambient gamma rays, and their suppression via active and passive shielding are well understood and reliably modeled. A similarly consistent picture was observed for the two cryogenic target detectors under different coincidence conditions with the veto systems. Some level of disagreement remains, however, which is believed to originate from limitations in the reliability of Geant4 at very low energies and is the subject of ongoing investigation. Meanwhile, the measured spectra in the target detectors also reveal a persistent LEE, which dominates the trigger rate in the CE$\nu$NS region of interest and significantly exceeds the expected signal (by about 4 orders of magnitude). The LEE remains the most critical background to be addressed and will require further investigation and mitigation in future phases of the experiment.

Looking ahead, the insights gained from this commissioning phase will guide the NUCLEUS experiment toward its next milestone: the technical run at the Chooz nuclear power plant. This run will serve to validate the full-scale experimental infrastructure in a reactor environment and to demonstrate the background levels achievable. These efforts will be complemented by continued R&D aimed at further improving detector sensitivity and refining the background model in the low-energy regime. Ultimately, the successful operation of NUCLEUS at Chooz will pave the way for the first detection of CE$\nu$NS from reactor antineutrinos using cryogenic detectors.

## ACKNOWLEDGMENTS

This work has been financed by the CEA, the INFN, the ÖAW, and partially supported by the TU Munich and the MPI für Physik. NUCLEUS members acknowledge additional funding by the DFG through the SFB1258 and the Excellence Cluster ORIGINS, by the European Commission through the ERC-StG2018-804228 "NUCLEUS," by the P2IO LabEx (ANR-10-LABX-0038) in the framework "Investissements d'Avenir" (ANR-11-IDEX-0003-01) managed by the Agence Nationale de la Recherche (ANR), France, by the Austrian Science Fund (FWF) through the "P 34778-N, ELOISE," and by Max-Planck-Institut für Kernphysik (MPIK), Germany.

## DATA AVAILABILITY

The data that support the findings of this article are not publicly available. The data are available from the authors upon reasonable request.

---

[1] D. Z. Freedman, Coherent neutrino nucleus scattering as a probe of the weak neutral current, Phys. Rev. D **9**, 1389 (1974).

[2] V. B. Kopeliovich and L. L. Frankfurt, Isotopic and chiral structure of neutral current, JETP Lett. **19**, 145 (1974), http://jetpletters.ru/ps/1776/article_27044.shtml.

[3] D. Akimov *et al.*, Observation of coherent elastic neutrino-nucleus scattering, Science **357**, 1123 (2017).

[4] D. Akimov *et al.*, First measurement of coherent elastic neutrino-nucleus scattering on argon, Phys. Rev. Lett. **126**, 012002 (2021).

[5] S. Adamski *et al.*, Evidence of coherent elastic neutrino-nucleus scattering with COHERENT's germanium array, Phys. Rev. Lett. **134**, 231801 (2025).

[6] V. De Romeri, O. G. Miranda, D. K. Papoulias, G. Sanchez Garcia, M. Tórtola, and J. W. F. Valle, Physics implications of a combined analysis of COHERENT CsI and LAr data, J. High Energy Phys. 04 (2023) 035.

[7] M. Cadeddu, F. Dordei, and C. Giunti, A view of coherent elastic neutrino-nucleus scattering, Europhys. Lett. **143**, 34001 (2023).

[8] B. C. Cañas, E. A. Garcés, O. G. Miranda, and A. Parada, Future perspectives for a weak mixing angle measurement in coherent elastic neutrino nucleus scattering experiments, Phys. Lett. B **784**, 159 (2018).

[9] M. Atzori Corona, M. Cadeddu, N. Cargioli, F. Dordei, and C. Giunti, Reactor antineutrinos CE$\nu$NS on germanium: CONUS+ and TEXONO as a new gateway to SM and BSM physics, Phys. Rev. D **112**, 015007 (2025).

[10] Luca Pattavina, Nahuel Ferreiro Iachellini, and Irene Tamborra, Neutrino observatory based on archaeological lead, Phys. Rev. D **102**, 063001 (2020).

[11] Caroline von Raesfeld and Patrick Huber, Use of CEvNS to monitor spent nuclear fuel, Phys. Rev. D **105**, 056002 (2022).

[12] A. Aguilar-Arevalo *et al.*, Search for coherent elastic neutrino-nucleus scattering at a nuclear reactor with CONNIE 2019 data, J. High Energy Phys. 05 (2022) 017.

[13] N. Ackermann *et al.*, Final CONUS results on coherent elastic neutrino-nucleus scattering at the Brokdorf reactor, Phys. Rev. Lett. **133**, 251802 (2024).

[14] I. Alekseev *et al.*, First results of the $\nu$GeN experiment on coherent elastic neutrino-nucleus scattering, Phys. Rev. D **106**, L051101 (2022).

[15] J. J. Choi *et al.*, Exploring coherent elastic neutrino-nucleus scattering using reactor electron antineutrinos in the NEON experiment, Eur. Phys. J. C **83**, 226 (2023).

[16] S. Karmakar *et al.*, New limits on the coherent neutrino-nucleus elastic scattering cross section at the Kuo-Sheng reactor-neutrino laboratory, Phys. Rev. Lett. **134**, 121802 (2025).

[17] D. Yu. Akimov *et al.*, Calibration and characterization of the RED-100 detector at the Kalinin nuclear power plant, J. Instrum. **19**, T11004 (2024).

[18] C. Cai *et al.*, Reactor neutrino liquid xenon coherent elastic scattering experiment, Phys. Rev. D **110**, 072011 (2024).

[19] C. Augier *et al.*, Fast neutron background characterization of the future Ricochet experiment at the ILL research nuclear reactor, Eur. Phys. J. C **83**, 20 (2023).

[20] G. Agnolet *et al.*, Background studies for the MINER coherent neutrino scattering reactor experiment, Nucl. Instrum. Methods Phys. Res., Sect. A **853**, 53 (2017).

[21] N. Ackermann *et al.*, CONUS+ experiment, Eur. Phys. J. C **84**, 1265 (2024); Eur. Phys. J. C **85**, 19(E) (2025).

[22] N. Ackermann *et al.*, Direct observation of coherent elastic antineutrino–nucleus scattering, Nature (London) **643**, 1229 (2025).

[23] R. Strauss *et al.*, The $\nu$-cleus experiment: A gram-scale fiducial-volume cryogenic detector for the first detection of coherent neutrino-nucleus scattering, Eur. Phys. J. C **77**, 506 (2017).

[24] R. Strauss *et al.*, Gram-scale cryogenic calorimeters for rare-event searches, Phys. Rev. D **96**, 022009 (2017).

[25] H. Abele *et al.*, Particle background characterization and prediction for the NUCLEUS reactor CE$\nu$NS experiment, arXiv:2509.03559.

[26] P. Adari *et al.*, EXCESS workshop: Descriptions of rising low-energy spectra, SciPost Phys. Proc. **9**, 001 (2022).

[27] G. Angloher *et al.*, Exploring CE$\nu$NS with NUCLEUS at the Chooz nuclear power plant, Eur. Phys. J. C **79**, 1018 (2019).

[28] H. de Kerret *et al.*, The Double Chooz antineutrino detectors, Eur. Phys. J. C **82**, 804 (2022).

[29] L. Perissé, A. Onillon, X. Mougeot, M. Vivier, T. Lasserre, A. Letourneau, D. Lhuillier, and G. Mention, Comprehensive revision of the summation method for the prediction of reactor $\nu e$ fluxes and spectra, Phys. Rev. C **108**, 055501 (2023).

[30] C. Goupy, Background mitigation strategy for the detection of coherent elastic scattering of reactor antineutrinos on nuclei with the NUCLEUS experiment, Ph.D. thesis, Université Paris Cité, 2024.

[31] A. Wex, Background suppression and cryogenic vibration decoupling for the coherent elastic neutrino scattering experiment NUCLEUS, Ph.D. thesis, Technische Universität München, 2025.

[32] G. Del Castello, LANTERN: A multichannel light calibration system for cryogenic detectors, Nucl. Instrum. Methods Phys. Res., Sect. A **1068**, 169728 (2024).

[33] A. Langenkämper *et al.*, A cryogenic detector characterization facility in the shallow underground laboratory at the Technical University of Munich, J. Low Temp. Phys. **193**, 860 (2018).

[34] Bluefors Oy, Ld dilution refrigerator measurement system https://bluefors.com/products/dilution-refrigerator-measurement-systems/ld-dilution-refrigerator-measurement-system/.

[35] A. Wex *et al.*, Decoupling pulse tube vibrations from a dry dilution refrigerator at milli-Kelvin temperatures, J. Instrum. **20**, P05022 (2025).

[36] https://www.aurubis.com/.

[37] V. Wagner *et al.*, Development of a compact muon veto for the Nucleus experiment, J. Instrum. **17**, T05020 (2022).

[38] A. Erhart *et al.*, A plastic scintillation muon veto for sub-Kelvin temperatures, Eur. Phys. J. C **84**, 70 (2024).

[39] AMPTEK, A250 charge sensitive preamplifier, https://www.amptek.com/internal-products/a250-charge-sensitive-preamplifier.

[40] B. Censier *et al.*, EDELWEISS read-out electronics and future prospects, J. Low Temp. Phys. **167**, 645 (2012).

[41] Axon' Group, Coaxial cables, https://www.axon-cable.com/en/coaxial-cables/.

[42] J. F. M. Rothe, Low-threshold cryogenic detectors for low-mass dark matter search and coherent neutrino scattering, Ph.D. thesis, Technische Universität München, 2021.

[43] SIS3316 16 CHANNEL VME DIGITIZER FAMILY, URL: https://www.struck.de/sis3316.html.

[44] CRYOCLUSTER Internal communication, The muon veto data acquisition software (unpublished) 2024.

[45] CRYOCLUSTER Internal communication, The versatile data acquisition software (vdaq3) (unpublished) 2025.

[46] C. S. Lamprecht, Meteostat phyton, https://github.com/meteostat/meteostat-python.

[47] C. Goupy, S. Marnieros, B. Mauri, C. Nones, and M. Vivier, Prototyping a high purity germanium cryogenic veto system for a bolometric detection experiment, Nucl. Instrum. Methods Phys. Res., Sect. A **1064**, 169383 (2024).

[48] X. Navick, Etude et optimisation de bolometres a mesure simultanee de l'ionisation et de la chaleur pour la recherche de la matiere noire, Ph.D. thesis, Université Paris Diderot—Paris 7, 1997.

[49] E. Gatti and P. F. Manfredi, Processing the signals from solid state detectors in elementary particle physics, Riv. Nuovo Cimento **9**, 1 (1986).

[50] S. Di Domizio, F. Orio, and M. Vignati, Lowering the energy threshold of large-mass bolometric detectors, J. Instrum. **6**, P02011 (2011).

[51] F. Pröbst, M. Frank, S. Cooper, P. Colling, D. Dummer, P. Ferger, G. Forster, A. Nucciotti, W. Seidel, and L. Stodolsky, Model for cryogenic particle detectors with superconducting phase transition thermometers, J. Low Temp. Phys. **100**, 69 (1995).

[52] H. Abele *et al.*, Sub-keV electron recoil calibration for cryogenic detectors using a novel X-ray fluorescence source, J. Low Temp. Phys. **221**, 265 (2025).

[53] M. Tinkham, *Introduction to Superconductivity (2nd Edition)* (Dover Publications, New York, 1996).

[54] C. Alduino *et al.*, Analysis techniques for the evaluation of the neutrinoless double-$\beta$ decay lifetime in $^{130}$Te with the CUORE-0 detector, Phys. Rev. C **93**, 045503 (2016).

[55] O. Azzolini *et al.*, Analysis of cryogenic calorimeters with light and heat read-out for double beta decay searches, Eur. Phys. J. C **78**, 734 (2018).

[56] Giorgio Del Castello *et al.*, Data analysis of the NUCLEUS experiment with the Diana framework, *Proc. Sci.*, TAUP2023 (2024) 257.

[57] F. Wagner, D. Bartolot, D. Rizvanovic, F. Reindl, J. Schieck, and W. Waltenberger, Cait: Analysis toolkit for cryogenic particle detectors in Python, Comput. Software Big Sci. **6**, 19 (2022).

[58] A. Thompson, D. Attwood *et al.*, X-ray data booklet, https://xdb.lbl.gov/xdb-new.pdf.

[59] L. Cardani, N. Casali, I. Colantoni, A. Cruciani, S. Di Domizio, M. Martinez, V. Pettinacci, G. Pettinari, and M. Vignati, Final results of CALDER: Kinetic inductance light detectors to search for rare events, Eur. Phys. J. C **81**, 636 (2021).

[60] G. Del Castello, Calibration and commissioning results of the NUCLEUS experiment, Ph.D. thesis, Rome University, 2025.

[61] H. Abele *et al.*, Observation of a nuclear recoil peak at the 100 eV scale induced by neutron capture, Phys. Rev. Lett. **130**, 211802 (2023).

[62] H. Abele *et al.*, The CRAB facility at the TU Wien TRIGA reactor: Status and related physics program, arXiv:2505.15227.

[63] S. R. Golwala, Exclusion limits on the WIMP nucleon elastic scattering cross-section from the cryogenic dark matter search, Ph.D. thesis, UC, Berkeley, 2000.

[64] M. Laubenstein, Screening of materials with high purity germanium detectors at the Laboratori Nazionali del Gran Sasso, Int. J. Mod. Phys. A **32**, 1743002 (2017).

[65] G. Angloher *et al.*, DoubleTES detectors to investigate the CRESST low energy background: Results from above-ground prototypes, Eur. Phys. J. C **84**, 1001 (2024); Eur. Phys. J. C **84**, 1227(E) (2024).

[66] R. Anthony-Petersen *et al.*, Low energy backgrounds and excess noise in a two-channel low-threshold calorimeter, Appl. Phys. Lett. **126**, 102601 (2025).

[67] L. Cardani *et al.*, High sensitivity phonon-mediated kinetic inductance detector with combined amplitude and phase read-out, Appl. Phys. Lett. **110**, 033504 (2017).

[68] C. L. Chang *et al.*, First limits on light dark matter interactions in a low threshold two channel athermal phonon detector from the TESSERACT Collaboration, arXiv:2503.03683.

[69] Ambra Mariani, Status and prospects of SABRE North, SciPost Phys. Proc. **12**, 026 (2023).

[70] R. Bernabei *et al.*, The DAMA project: Achievements, implications and perspectives, Prog. Part. Nucl. Phys. **114**, 103810 (2020).

[71] J. D. Hakenmüller, Looking for coherent elastic neutrino nucleus scattering with the CONUS experiment, Ph.D. thesis, Heidelberg University, Germany, 2020.

[72] M. Kaznacheeva, Exploring the sub-keV energy region with the CRESST and NUCLEUS experiments, Ph.D. thesis, Technische Universität München, 2024.